\documentclass[%
preprint,
amsmath,amssymb,
aps,
pra,
]{revtex4-2}

\usepackage{graphicx}
\usepackage{amsmath}
\usepackage{multirow}
\usepackage{color}
\usepackage{float}

\begin{document}

\author{Lorenz S. Cederbaum}
\email{Lorenz.Cederbaum@pci.uni-heidelberg.de}
\affiliation{Theoretische Chemie, Physikalisch-Chemisches Institut, Universit\"at Heidelberg, Im Neuenheimer Feld 229, Heidelberg D-69120, Germany}
\author{Jacqueline Fedyk}
\affiliation{Theoretische Chemie, Physikalisch-Chemisches Institut, Universit\"at Heidelberg, Im Neuenheimer Feld 229, Heidelberg D-69120, Germany}

\title{Activating Cavity by Electrons}


\begin{abstract}
 
The interaction of atoms and molecules with quantum light as realized in cavities has become a highly topical and fast growing field of research. This interaction leads to the formation of hybrid light-matter states giving rise to new phenomena and opening up new pathways to control and to manipulate properties of the matter. In this work, we substantially extend the scope of the interaction by allowing free electrons to enter the cavity and merge and unify the two active fields of electron scattering and quantum-light-matter interaction. In the presence of matter, a new kind of metastable states are formed which exist at a large range of electron energies. The properties of these states depend strongly on the frequency and on the light-matter coupling of the cavity. The incoming electrons can be captured by the matter inside the cavity solely due to the presence of the cavity. The findings are substantiated by an explicit example and general striking consequences are discussed.

\end{abstract}





\maketitle
 
\section*{Quantum light and electron scattering} \label{sec:Introduction}

The energy levels of atoms and molecules change dramatically in confined spaces like that inside a cavity due to the coupling of matter excitations with the quantized radiation field. By this coupling hybrid light-matter states are formed opening up new pathways to manipulate and even control static and dynamic properties of the matter. Much work has been published reporting many possibilities to amplify or suppress available mechanisms and even to induce new ones by the presence of the quantized radiation field. Among the many examples reported, we mention here the possibilities to enhance energy transfer \cite{Cavity_ET_1,Cavity_CT_3} and charge transfer \cite{Cavity_CT_1,Cavity_CT_2,Cavity_CT_3,Cavity_CT_4}, to strongly vary the rate of spontaneous emission \cite{Cavity_Varying_Spontaneous_Emission}, to enhance or to completely suppress interatomic Coulombic decay \cite{Cavity_ICD}, to control photochemical reactivity \cite{Cavity_Chem_Reac_1,Cavity_Chem_Reac_2,Cavity_Suppressing_Reactions_Feist}, to control chemical reactions by varying the quantized field  \cite{Cavity_Chem_Reac_3,Cavity_Chem_Reac_4,Cavity_Chem_Reac_5,Cavity_Chem_Reac_6,Cavity_GS_Reactivity}, and to induce new molecular non-adiabatic processes not available in free space \cite{Cavity_Chem_Reac_5,Cavity_Non_Adiab_1,Cavity_Non_Adiab_2,Cavity_LICI_1,Cavity_LICI_3,Cavity_LICI_4,LICI_LiF_Huo,Cavity_Coll_CI}.

Incoming electrons are known to interact strongly with atoms and molecules in particular at kinetic energies at which the electrons temporarily attach to them forming so called resonances or metastable states \cite{Resonances_Atoms_Schulz,Resonances_Molecules_Schulz,Resonances_Review_Domcke}. The palette of phenomena due to resonances is very large, has been widely studied and is still an active field of research. The resonances show up in electron scattering experiments as enhancements of the scattering cross section at characteristic energies with widths typical for the lifetime of the resonance in question. In the case of molecules, many kinds of reactions take place of which we would like to mention only a few. May be the most important reaction is by a process called dissociative electron attachment \cite{Resonances_ACS_Symposium,Resonances_Review_Domcke,Resonances_DEA_Simons}, where an incoming electron first forms a resonance state with the molecule leading to the fragmentation of the molecule and the production of an anion as one of the fragments. This mechanism has been shown to play an important role in the atmosphere and in strand breaking of DNA in radiation damage \cite{Resonances_DEA_Sanche_1,Resonances_DEA_Sanche_2}. A recent example is the production of molecular oxygen from carbon dioxide \cite{Resonances_DEA_CO2}. In another kind of reactions involving an electron and a molecule in a metastable compound, called bond breaking by a catalytic electron, the formed resonance dissociates into nonradical neutral molecular subunits and a free electron, which plays the role of a catalyst \cite{Catalytic_electron_1,Catalytic_electron_2,Catalytic_electron_3}. In the resonance, the activation energy of a chemical reaction can be lowered enabling new reactions. An example is the barrierless path for the cis–trans isomerization of maleonitrile formed on the resonance surface where the electron leaves the system after yielding the desired isomer \cite{Resonance_Barrierless_Isomerization}. 

We shall demonstrate that the two areas of research, matter in quantum light and resonant electron scattering, can be merged and unified giving rise to new fundamental and intriguing phenomena. As one may anticipate that the resonance phenomena known in the absence of quantum light will be modified by the presence of this light, we concentrate in this first work on the appearance of a new type of resonances which is only present in quantum light. 

To easily distinguish between atomic and molecular resonances known in the absence of quantum light and resonances which may appear due to the presence of the quantum light, we would like to call the former {\it{natural resonances}} and the latter {\it{quantum light-induced resonances}}. Whenever unambiguous, the name {\it{light-induced resonances}} is used for brevity. 

\section*{Setup and application} \label{sec:Results_and_Discussion}

\begin{figure}[H]
	\begin{center}
		\includegraphics[width=10cm]{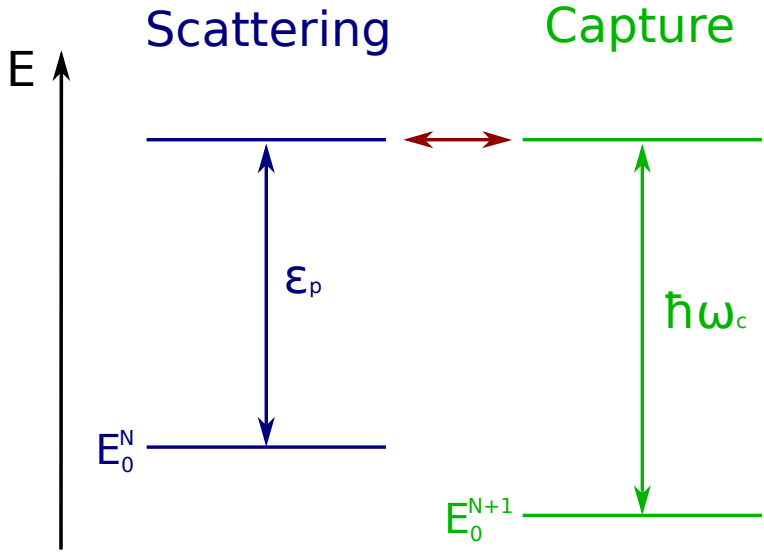}
		\caption{Schematic picture of the scenario investigated. An impinging electron of kinetic energy $\epsilon_p$ can scatter off a target atom or molecule which is in its ground state of energy $E^{N}_0$ and possesses $N$ electrons. As is the case for most atoms and molecules, the target can bind an electron and become a negative ion (anion) of energy $E^{N+1}_0$. We shall show that in a cavity of frequency $\omega_c$, the hybrid state formed by the anion and a cavity photon becomes a resonance leading to dramatic features in the scattering cross section of the impinging electron. Moreover, due to the cavity the impinging electron can be captured by the target.} 
		\label{fig:system}
	\end{center}
\end{figure}

We consider an incoming electron of kinetic energy $\epsilon_p$ impinging on a target atom or molecule. If this target system does not possess a natural resonance in the vicinity of $\epsilon_p$, the central question arises of how the quantum light can at all induce a resonance. For the ease of discussion we assume, without loss of generality, that the target system is neutral. The scenario we address is depicted schematically in Fig.~\ref{fig:system}. We shall show below that the hybrid state formed by the anion of the target system and a cavity photon becomes a resonance and mention that nearly all atoms \cite{Negative_Ions_Atoms_1} and the bulk of molecules \cite{Negative_Ions_Book_Herzberg,Multiply_Charged_Anions_1} possess a stable negative ion. We shall see that impinging electrons of suitable kinetic energy populate this resonance and thus activate the cavity, i.e., induce a cavity photon. 

We consider an electronic system of $N+1$ electrons in a cavity with a quantized light mode (cavity mode) of frequency $\omega_c$ and polarization direction $\vec{e}_c$. The matter-cavity Hamiltonian reads \cite{Cohen_Tannoudji_Book,Feist_PhysRevX}:

\begin{align}\label{Matter-Cavity-Hamiltonian}
	H = H_e + \hbar\omega_c\hat{a}^\dagger \hat{a} + g_0 \vec{e}_c\cdot \vec{d}(\hat{a}^\dagger + \hat{a}),
\end{align}

where $H_e$ is the electronic Hamiltonian of the system, $\vec{d}$ is the dipole operator of the system and $g_0$ is the coupling strength of the cavity. The quadratic dipole self-energy term is neglected as it is only of relevance for very strong coupling. One should be aware that the eigenstates of $H_e$ include both bound and electron scattering states.

Initially, the target atom or molecule is in its ground state $\Phi^{N}_{0}$ of energy $E^{N}_0$, the incoming electron of momentum $p$ has the kinetic energy $\epsilon_p=p^2/2m$, and the cavity is at rest, i.e., it has zero photons. We shall denote this initial state as $|I,p,0_c\rangle = |\Phi^{N+1}(p),0_c\rangle$ and its energy is $E_I = E^{N}_0 + \epsilon_p$. The situation is schematically depicted on the left hand side of Fig.\ref{fig:system}. Other scattering states are analogously denoted by $|\Phi^{N+1}(k),0_c\rangle$. In the above Hamiltonian, the matter-cavity interaction term $H_{mc} = g_0 \vec{e}_c\cdot \vec{d}(\hat{a}^\dagger + \hat{a})$ couples states of the same total number of electrons, but with a different number of cavity photons. As a result, a scattering state $|\Phi^{N+1}(k),0_c\rangle$ can couple to a completely different type of state where the $N+1$ electrons form a bound state $\Phi^{N+1}_{0}$ of energy $E^{N+1}_0$ and the cavity has one photon. This state is denoted by $|\Phi^{N+1}_{0},1_c\rangle$ and its energy is $E^{N+1}_0 + \hbar\omega_c$. The quantity $E^{N}_0-E^{N+1}_0$ is the electron affinity of the target. See the right hand side of Fig.~\ref{fig:system}.        

To describe the scattering process and the formation of the matter-cavity hybrid resonance, it is convenient to compute the matrix elements of the so-called T-matrix \cite{Davydov_Book}. In our situation the T-matrix reads 

\begin{align}\label{T_Matrix}
T = H_{mc} + H_{mc}(E_I - H_e - \hbar\omega_c\hat{a}^\dagger \hat{a} +i0^+)^{-1}T,
\end{align}

where $0^+$ is a positive infinitesimal. In the space of the scattering states $|\Phi^{N+1}(k),0_c\rangle$ and the bound hybrid state $|\Phi^{N+1}_{0},1_c\rangle$, the T-matrix for the resonant scattering can be computed in closed form as shown in the Supplementary Materials. Relevant are two kinds of matrix elements of the T-matrix. The first is between the initial state $|I,p,0_c\rangle$ introduced above and the final state $|F,p',0_c\rangle = |\Phi^{N+1}(p'),0_c\rangle$ of the scattering process and determines the resonant scattering cross section for an incoming electron of momentum $p$ and an outgoing electron of momentum $p'$. The second is between the initial state $|I,p,0_c\rangle$ and the hybrid state $|\Phi^{N+1}_{0},1_c\rangle$ and is related to the processes of capturing the electron by the target. We first concentrate on the scattering and return later to the capture.

The relevant result for resonant scattering takes on the appearance:
 
\begin{align}\label{Scattering_T_Matrix_Element}
|\langle F,p',0_c|T|I,p,0_c\rangle|^2 = |\gamma_{p'0}|^2 \bigg|\frac{1}{\epsilon_p - \epsilon_c - F(\epsilon_p)}\bigg|^2|\gamma_{p0}|^2.
\end{align}

This appearance reminds of that of a discrete electronic state of energy $\epsilon_c$ embedded in the continuum and interacting with it \cite{Resonances_Review_Domcke}. In a cavity, the discrete state's energy is explicitly $\epsilon_c=E^{N+1}_0-E^{N}_0+\hbar\omega_c$ and describes the electron affinity of the target shifted into the continuum by a $\hbar\omega_c$ photon of the cavity, see Fig.~\ref{fig:system}. The interaction with the continuum is via the the matter-cavity interaction term $H_{mc}$ and turns the discrete state into a hybrid matter-cavity resonance with a finite lifetime. This interaction gives rise to a complex and energy dependent level shift

\begin{align}\label{Complex_Level_Shift}
F(\epsilon_p) = \Delta(\epsilon_p) - \frac{i}{2}\Gamma(\epsilon_p) = \sum_{k} \frac{|\gamma_{k0}|^2}{\epsilon_p - \epsilon_k +i0^+}.
\end{align}

The quantity $\Delta$ quantifies the energy shift caused by the interaction matrix elements  $\gamma_{k0}=\langle\Phi^{N+1}(k),0_c|H_{mc}|\Phi^{N+1}_{0},1_c\rangle $ and $\Gamma$ determines the lifetime $\tau=\hbar/\Gamma$ of the created resonance. We mention that the real shift function $\Delta$ follows from the knowledge of the width function $\Gamma$, see Ref.~\cite{Resonances_Review_Domcke} and Supplementary Materials. If the dependence of $\Gamma$ on the energy can be neglected, the above expression (\ref{Scattering_T_Matrix_Element}) takes on the well known form of a Breit-Wigner cross section \cite{Davydov_Book,Taylor_Book}.

\subsection*{Show-case example} \label{sec:Show_case_example}

Although the true potential of the new idea proposed here is for molecules, we would like in this first work to accompany the general theory by an explicit and transparent show-case example to illustrate and quantify the findings. We chose the H atom as the target. One-electron potentials are often used to describe the electron attaching to it and forming the H$^-$ anion, see Ref.~\cite{Model_Potential_H_anion} and references therein. We adopt here the three dimensional potential (in atomic units) employed to compute the photodetachment of H$^-$ by strong high-frequency linearly polarized laser pulses \cite{Model_Potential_H_anion}

\begin{align}\label{Model_Potential}
V(r) = -V_0 e^{-r^2/r_0^2},
\end{align}  

where $V_0= 0.3831087$ and $r_0=2.5026$. This potential supports a single bound state and reproduces its energy and  s-wave scattering length in full agreement with accurate variational results. Using this potential, the bound state of s symmetry and continuum states of p symmetry have been computed numerically within radial boxes of varying size $L$, and with them the coupling elements $\gamma_{k0}=\langle\Phi^{N+1}(k),0_c|H_{mc}|\Phi^{N+1}_{0},1_c\rangle $ have been determined. These elements then served to calculate the shift and width functions $\Delta$ and $\Gamma$. The convergence with respect to the size of the box has been checked. The calculations are described in the Supplementary Materials.

 \begin{figure}[H]
 	\begin{center}
 		\includegraphics[width=10cm]{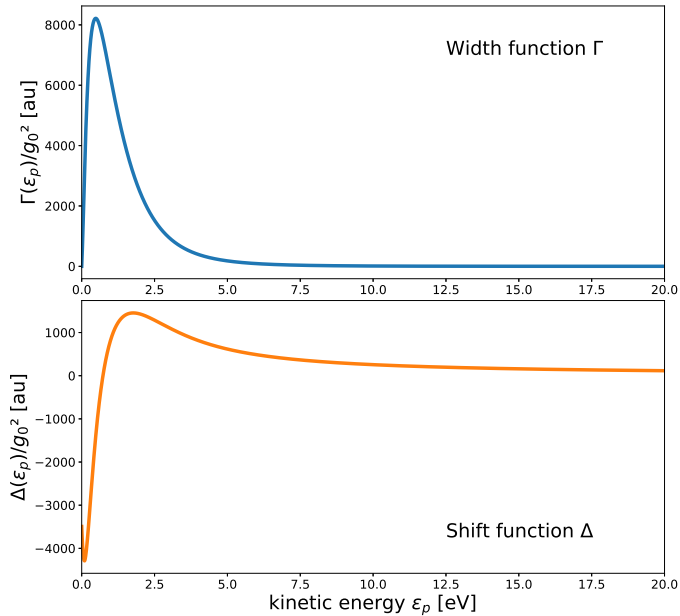}
 		\caption{The shift and width functions $\Delta$ and $\Gamma$ for resonant electron scattering from the H atom in a cavity. The model potential in Eq.~(\ref{Model_Potential}) has been applied. $\epsilon_p$ is the kinetic energy of the impinging electron. $\Delta$ and $\Gamma$ account for the energy shift and decay rate the discrete state depicted on the r.h.s. of Fig.~\ref{fig:system} experiences due to its interaction with the continuum. Note that the plots show the quantities $\Delta/g^2_0$ and $\Gamma/g^2_0$ which are independent of the coupling strength $g_0$ of the cavity.}
 		\label{fig:decay_width_delta}
 	\end{center}
 \end{figure} 

The computed shift and width functions $\Delta$ and $\Gamma$ are shown in Fig.~\ref{fig:decay_width_delta}. Inspection of Eq.~(\ref{Complex_Level_Shift}) shows that these quantities characterizing the impact of the continuum on the discrete state, depend quadratically on the coupling strength $g_0$ of the cavity. While $\Gamma$ is, except at threshold  $\epsilon_p=0$, a positive function, $\Delta$ can take on positive as well as negative values. This implies that the energy $\epsilon_c$ of the discrete state can be shifted to higher or lower energies by its interaction with the continuum depending on the value of the photon energy $\hbar\omega_c$, see the denominator in Eq.~(\ref{Scattering_T_Matrix_Element}) and text below. 

It is common to define the energy of the resonance, $E_{res}$, at the energetic position where the real part of the denominator of the T-matrix vanishes \cite{Resonances_Review_Domcke}. In our case of a quantum light-induced resonance, this definition leads to finding the kinetic energy where $\epsilon_p = \epsilon_c + \Delta(\epsilon_p)$ is fulfilled. In the specific example of the H atom, $\Delta(\epsilon_p)$ is seen in Fig.~\ref{fig:decay_width_delta} to vanish at about $\epsilon_p=0.73$~eV where it changes its sign from being negative at lower and positive at higher values of the energy. Accordingly, the resonance is shifted to lower energies than $\epsilon_c$ if $\hbar\omega_c$ is chosen to be smaller than $(0.75 + 0.73)=1.48$~eV  and to higher energies otherwise. An interesting situation arises if $\hbar\omega_c<1.48$~eV and the cavity coupling constant $g_0$ is large. Then, the discrete state can be shifted by the continuum below threshold and become a bound state of an anion and a cavity photon. For more details, see Supplementary Materials, in particular Fig.~S5. We remind that all of these is controlled by the cavity in contrast to the situation in natural resonances where the resonance is an entity given by nature.

\subsection*{Cross sections} \label{sec:cross_sections}

To obtain the integral cross section for resonant scattering, one has to integrate the expression (\ref{Scattering_T_Matrix_Element}) over the final angles of the outgoing electron and average over the molecular orientations. In case a single outgoing wave dominates the scattering, the cross section takes on the simple appearance \cite{Resonance_SCattering_N2_1}

\begin{align}\label{Scattering_Cross-Section}
\sigma = \frac{2\pi}{\epsilon_p} \Gamma(\epsilon_{p'})\bigg|\frac{1}{\epsilon_p - \epsilon_c - F(\epsilon_p)}\bigg|^2 \Gamma(\epsilon_{p}).
\end{align}

In our explicit example of resonant scattering from H, the scattering is elastic, i.e., $\epsilon_{p'} = \epsilon_{p}$. We have kept $\Gamma(\epsilon_{p'})$ for later purposes.  

\begin{figure}[H]
	\begin{center}
		\includegraphics[width=10cm]{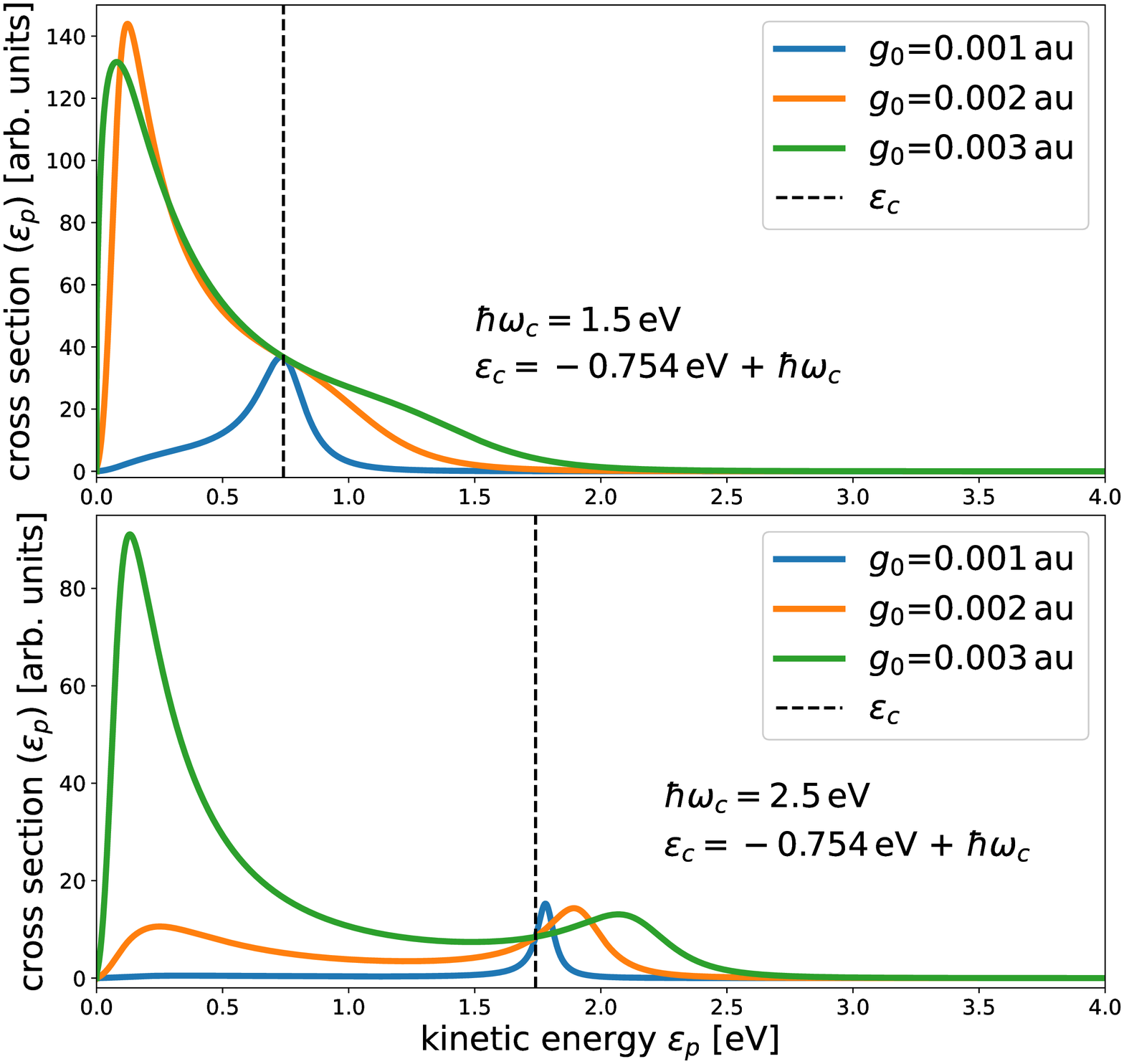}
		\caption{Illustrative examples of the electron scattering cross section for H atom in a cavity. The employed width and shift functions are depicted in Fig.~\ref{fig:decay_width_delta}. Shown are cross sections for several values of the coupling strength $g_0$ and photon energy $\hbar\omega_c$ of the cavity. The cross sections are seen to strongly depend on these quantities. $\epsilon_c=E^{N+1}_0-E^{N}_0+\hbar\omega_c$ is the energy of the discrete state above the threshold of the continuum, see Fig.~\ref{fig:system}. This energy is indicated by a broken vertical line and allows one to better see the impact of the interaction of the discrete state with the continuum turning it into a resonance. The electron affinity of the H atom is 0.754 eV.} 
		\label{fig:cross_section}
	\end{center}
\end{figure}

Using the shift and width functions shown in Fig.~\ref{fig:decay_width_delta}, the respective resonant cross sections are depicted in Fig.~\ref{fig:cross_section} for three values of the cavity coupling strength $g_0$ and two values of the cavity frequency. Due to the dependence of the shift and width functions on the energy, the cross sections are seen to strongly differ from the typical Lorentzian Breit-Wigner form. The cross sections are sensitive to the cavity coupling strength as well as to the cavity frequency. Even for the weakest coupling strength, the structure of the cross section is asymmetric for $\hbar\omega_c=1.5$~eV, and becomes Breit-Wigner-like only at higher cavity frequency. For the intermediate and larger values of the coupling strength, it is eye catching that the cross section is not localized energetically around a specific energy, but is rather spread over a broad energy interval. For $\hbar\omega_c=1.5$~eV, at which the shift function nearly vanishes (see text above and Fig.~\ref{fig:decay_width_delta}), the cross sections are seen to peak close to threshold and much less at the energy of the expected resonance at which the cross section for the weakest coupling strength peaks.  The situation is even more pronounced for the larger cavity frequency $\hbar\omega_c=2.5$~eV, where a clear double-peaked cross section is found, one peak at the energy expected for the resonance on the ground of the weak coupling case, and one peak closer to threshold making the cross section to extend over a large energy range. 

The found peculiar behavior of the cross section can be understood by examining expression (\ref{Scattering_Cross-Section}) for resonant scattering. The denominator is $[(\epsilon_p - \epsilon_c - \Delta(\epsilon_p)]^2 +  [\Gamma(\epsilon_p)/2]^2$. For small coupling strength $g_0$, the cross section is seen to peak around the resonance position  $E_{res}$, i.e., where $E_{res} = \epsilon_c + \Delta(E_{res})$ is fulfilled. The shift and width functions grow quadratically with $g_0$ and, therefore, as the coupling strength grows $[(\epsilon_p - \epsilon_c - \Delta(\epsilon_p)]^2$ can become small at several and even for a range of values of kinetic energies $\epsilon_p$. See Figs.~S3 and S4 and the accompanying text in the Supplementary Materials. Depending on the structure of the width function, the resonant cross section can become substantial also at these values of kinetic energies.
  

\subsection*{Electron capture probabilities} \label{sec:electron_capture}

We now turn to electron capture. Without a cavity an isolated atom can only capture an electron by emitting the excess energy as a photon, a process called photo-recombination widely studied in media with
low atomic densities \cite{Sobelman,Photorecombination_Review_Mueller}. The capture cross section by photo-recombination is rather small. The capture cross section has been shown to be substantially enhanced by the presence of neighboring atoms and molecules via a process called interatomic Coulombic electron capture \cite{ICEC_JPB,ICEC_First_ab_initio}. Can the presence of a cavity make electron capture by a single isolated atom or molecule an efficient process?  

To answer this basic question we turn to the total scattering state $|\Psi^+_p\rangle$. This state fulfills the well known Lippmann-Schwinger equation \cite{Gottfried_Book,Taylor_Book} which is closely related to the T-matrix equation (\ref{T_Matrix}). When calculating the T-matrix, one also obtains the total scattering state. While the projection of the final scattering states $|F,p',0_c\rangle$ introduced above on the total scattering state $|\Psi^+_p\rangle$, i.e., $|\langle F,p',0_c|\Psi^+_p\rangle|^2$, describes the scattering process, the projection of the discrete state $|\Phi^{N+1}_{0},1_c\rangle$ on the total scattering state, i.e., $|\langle \Phi^{N+1}_{0},1_c |\Psi^+_p\rangle|^2$, describes the capture of the impinging electron by the target. This quantity is calculated in the Supplementary Materials. To assess the importance of the capture and its dependence on the energy, we consider the probability to capture an electron with kinetic energies in a small energy interval $\delta_{\epsilon_p}$ around $\epsilon_p$. As shown in the Supplementary Materials, the capture probability takes on the appearance 

\begin{align}\label{Capture_Probability}
	P_{capture} = \dfrac{1}{2 \pi} \dfrac{\Gamma(\epsilon_p) \delta_{\epsilon_p}}{(\epsilon_p - \epsilon_c - \Delta(\varepsilon_p))^2 + (\Gamma(\epsilon_p) + \Gamma_{phot})^2/4}.
\end{align}   

\noindent The new quantity $\Gamma_{phot}$ will be discussed below. 

\begin{figure}[H]
	\begin{center}
		\includegraphics[width=10cm]{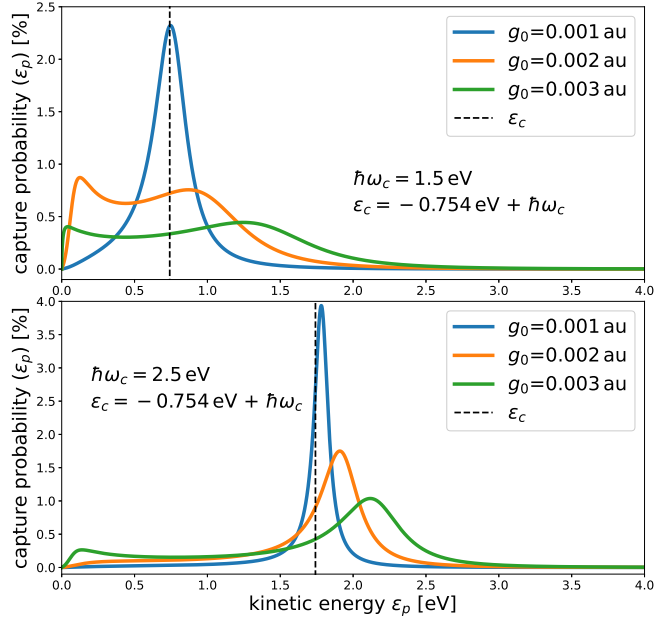}
		\caption{Illustrative examples of the electron capture probability by the H atom in a cavity. Shown are capture probabilities for several values of the coupling strength $g_0$ and photon energy $\hbar\omega_c$ of the cavity. The corresponding scattering cross sections are depicted in Fig.~\ref{fig:cross_section}. The lifetime of the cavity has been assumed to be 20 fs, i.e., $\Gamma_{photon}=33$~meV, and the probability relates to electrons in a kinetic energy interval of $\delta_{\epsilon_p}=0.01$~eV. The energetic position of the discrete state embedded in and interacting with the continuum is drawn as a broken vertical line.} 
		\label{fig:capture_probability}
	\end{center}
\end{figure}

The capture process needs some further discussion. The discrete state is, of course, not an eigenstate of the Hamiltonian and it decays into the continuum as discussed above. However, its projection on the total scattering state $|\Psi^+_p\rangle$ has to be understood as an average value of the occupation of this discrete state and will be found if measurements are made. But there is more to it. The discrete state is a hybrid cavity-matter state and it can also decay by emitting the cavity photon and become a truly bound anion which cannot decay anymore. Indeed, quantum light as realized in nowadays cavities is subject to the finite lifetime of the cavity photon, see, e.g., Ref.~\cite{Cavity_Dynamics_Polaritons_Sfeir}. This decay of the cavity photon can be taken into account \cite{Cavity_Losses_Feist} by employing Lindblad master equations or, equivalently, by solving the Schr\"odinger equation for the non-Hermitian Hamiltonian obtained by augmenting the cavity Hamiltonian $\hbar\omega_c\hat{a}^\dagger   \hat{a} \rightarrow \hbar\omega_c\hat{a}^\dagger \hat{a} -i\Gamma_{phot}/2$. Inspection of the T-matrix (\ref{T_Matrix}), shows how the cavity photon decay rate $\Gamma_{phot}$ enters the denominator, giving rise to the overall width $(\Gamma(\epsilon_p) + \Gamma_{phot})$ of the resonance.


To illustrate the capture of the impinging electron by the H atom, the capture probability $P_{capture}$ is shown in Fig.~\ref{fig:capture_probability} as a function of energy for the same values of the cavity frequencies and cavity coupling strengths as used for the scattering cross sections in Fig.~\ref{fig:cross_section}. The photon decay rate $\Gamma_{phot} = 33$~meV used corresponds to a cavity photon lifetime of $20$~fs discussed for a nano cavity \cite{Cavity_Dynamics_Polaritons_Sfeir}. The capture probability is seen to vary strongly with the cavity frequency and with the coupling strength. Interestingly, the capture is most efficient for the weaker coupling strength. This is related to the fact that the decay width of the resonance $\Gamma(\epsilon_p)$ scales quadratically with the coupling strength, see also Fig.~\ref{fig:decay_width_delta}. Consequently, the decay of the discrete state into the electron continuum can be made slow for weak coupling and this, in turn, gives rise to a longer lifetime of the resonance and with it to a higher capture probability.

\section*{Future prospects} \label{sec:Conclusions_and-Outlook}

Normally, hybrid states are populated by external light by addressing a polariton. Recently, the possibility to pump the cavity mode 
with a laser pulse has also been explored computationally \cite{Cavity_Mode_Pumped_Feist,Cavity_Mode_Pumped_Agnes}. In this work we have shown that the cavity can be activated by electrons. This opens the door to a completely new kind of physics and chemistry. A hybrid resonance state consisting of an electron attached to the target and a cavity photon is formed and its decaying properties are discussed. This light-induced resonance can decay into the electron continuum and also by loss of a cavity photon thus creating a stable atom or molecule with an extra electron. As the cavity frequency and coupling strength determine the energetic position and decay properties of the light-induced resonance, one can have full control of the resonance and its properties in stark contrast to free atoms and molecules where the  resonances and their properties are given by nature. 

\begin{figure}[H]
	\begin{center}
		\includegraphics[width=9cm]{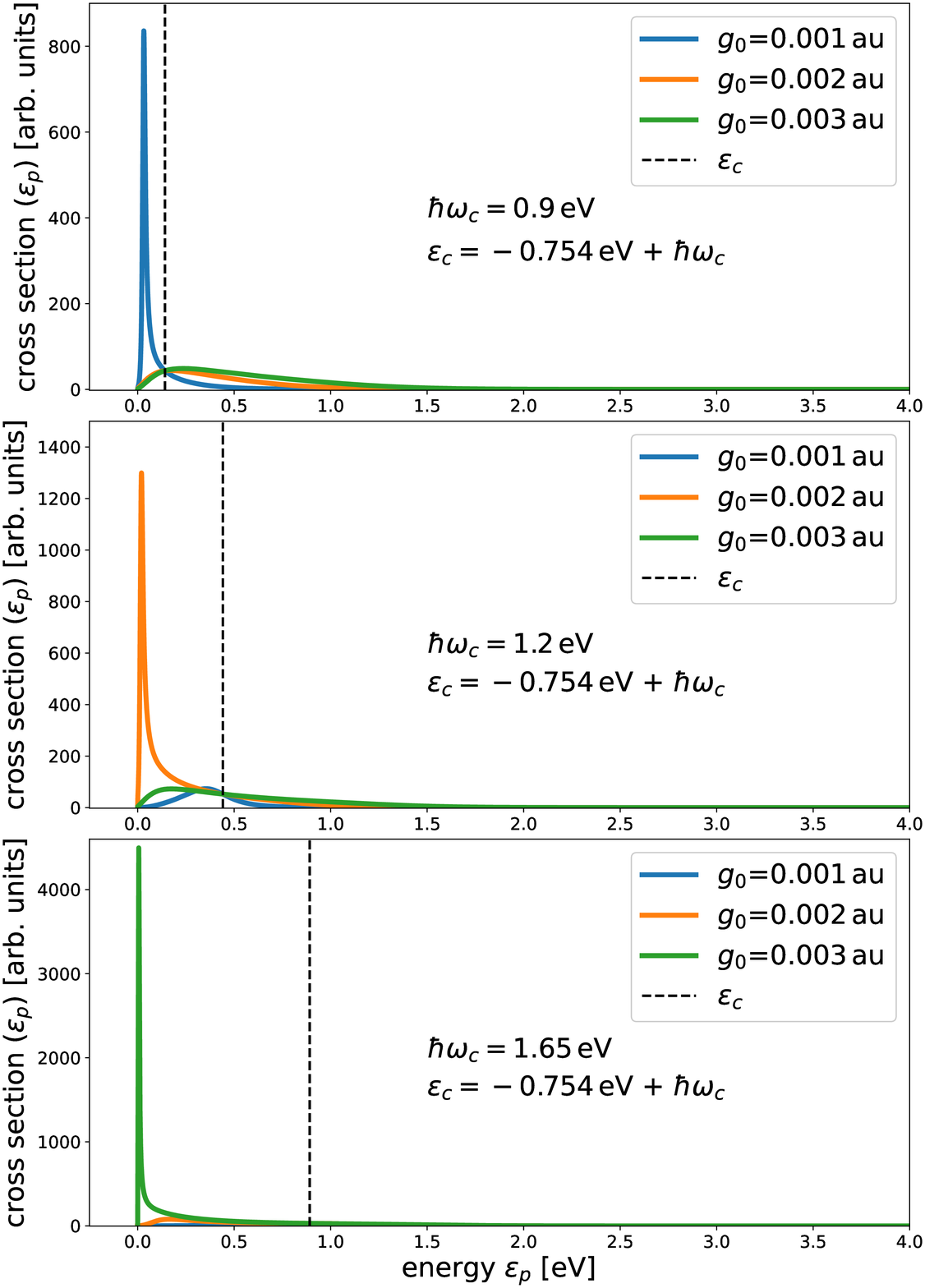}
		\caption{Illustrative examples of cross sections peaking close to threshold by controlling the cavity frequency for different coupling strengths. The energetic position of the discrete state embedded in and interacting with the continuum is drawn as a broken vertical line. Its energy depends linearly on the cavity frequency. This discrete state is shifted downwards in energy due to its interaction with the continuum and becomes a resonance due to its coupling to the continuum. As the shift depends quadratically on the coupling strength $g_0$, one can by choosing the photon energy $\hbar\omega_c$ control the location of the resonance. A resonance close to threshold leads to a high and peaked cross section. The photon energy $\hbar\omega_c=0.9$~eV brings the resonance for $g_0=0.001$~au close to threshold where it dominates the cross sections of the larger coupling strengths. In analogy, the other values of the photon energy shown are chosen to bring the resonance close to threshold for the other values of the coupling strength. See also Fig.~S6 in Supplementary Materials, demonstrating the sensitivity to the cavity photon lifetime for resonances at threshold.} 
		\label{fig:cross_section_threshold}
	\end{center}
\end{figure}

There is a vast number of applications and phenomena to be expected. A few directions are briefly discussed in the following. Cavities of different kinds are available, like optical and plasmonic nanocavities, and over the last decade, there has been a tremendous progress of quantum cavity technologies, for a review, see, e.g., Ref.~\cite{Cavity_Review_Technology}. Depending on the properties of the cavities one can choose suitable scenarios to investigate. For instance, if the cavity frequency available is small, systems with small electron affinities in combination with low electron kinetic energies can be investigated. In general, there are highly interesting phenomena near threshold if a low lying resonance is encountered \cite{Resonances_Review_Domcke}, and in cavity we are rather flexible to place the resonance at will. Let us return to the example of the H atom and choose the cavity frequency such that for a given coupling strength the position of the emerging resonance is close to threshold. The calculated cross sections are collected in Fig.~\ref{fig:cross_section_threshold} for three values of the coupling strength. A resonance close to threshold leads to a high and peaked cross section. In the upper panel the photon energy $\hbar\omega_c=0.9$~eV creates a light-induced resonance close to threshold for the weakest coupling strength. As seen in the figure, the respective cross section strongly dominates those for the stronger coupling values. It is illuminating to notice that the resonance position follows from the equation $E_{res}=\epsilon_c + \Delta(E_{res})$ which for $\hbar\omega_c=0.9$~eV has solutions at negative energies for the stronger coupling strengths, i.e., instead of a resonance, a bound hybrid state appears below threshold. The impact of these bound states on the respective cross sections is much weaker than that of a resonance close to threshold in spite of the larger coupling strengths. The other panels of Fig.~\ref{fig:cross_section_threshold} demonstrate that one can selectively enhance the cross section for a desired coupling strength. The visualization of the latter equation, a figure showing the respective capture probabilities and additional cross sections are collected in the Supplementary Materials. A comparison of Fig.~\ref{fig:cross_section_threshold} which is in the absence of cavity losses and of Fig.~S6 in the Supplementary Materials which takes account of losses, strikingly underline that the measured cross sections sensitively reflect the cavity photon lifetime for resonances at threshold. This important observation may provide a tool to investigate details of cavity losses. 

Anions of most atoms and of many molecules exist only in their electronic ground state, but there are interesting systems, like fullerenes, which possess several excited bound anionic states, see, e.g., Ref.~\cite{Anions_C20_YiFan}. Accordingly, there can be several light-induced hybrid resonances when in cavity. These resonances may be close by each other and even overlap within their decay width functions giving rise to interference effects in the cross sections and in the capture probabilities. If they do not overlap, one can in principle populate only one of them selectively and when the cavity photon decays, one is left in a single selected excited state of the anion. 

Of course, the atom or molecule in the cavity can also be a positive ion which is neutralized by the electron capture in the cavity. As the neutral system possess infinitely many Rydberg states close to the ionization threshold, there is an infinite manifold of discrete hybrid states in quantum light made of a Rydberg state and a cavity photon. All these states interact with the electron continuum giving rise to a complex light-induced resonance structure which makes its calculation rather intricate. The resulting electron capture process should take place already at very small cavity photon energies as the Rydberg series converges towards the ionization threshold. Due to the dense Rydberg manifold, we expect the capture to be efficient. 

An important issue is the interplay of natural and light-induced resonances. If the energies of these two kinds of resonances are within their width functions, they will interact and severe interference effects are to be expected. As one can vary the photon energy, it is possible in many cases to bring the light-induced resonance to the energetic vicinity of the natural resonance and in this way have an impact on the effect of the latter. For example, as discussed in the introduction, molecular resonances can lead to the dissociation of the molecule by dissociative electron attachment, which is a highly important process triggered by a resonance. By creating a light-induced resonance in the vicinity of the natural resonance, one can, in principle, impede the dissociation. The interplay between the resonances will depend on the cavity photon energy and the cavity coupling strength which can be varied to better control the outcome.      

Molecules possess internal degrees of freedom, in particular vibrations. The resonant scattering in free molecules leads to severe vibrational excitations and this is expected to also be the case in cavity with the light-induced resonance. The formalism employed here for cavity can be extended to take account of vibrations following the same line as done for free molecules by replacing $\epsilon_p$ in the complex shift function $F(\epsilon_p)$ by $\epsilon_p - H_v$, where $H_v$ is the vibrational Hamiltonian of the target \cite{Resonances_Review_Domcke,Local_vs_Nonlocal_1981}. As a consequence, this basic function becomes an operator in nuclear coordinate space playing the role of an energy dependent and nonlocal potential for the nuclear motion in the resonance which has an impact on observations even far from threshold \cite{Resonance_SCattering_N2_1}. Close to threshold there is a plethora of strong effects due to the nonadiabaticity imposed by this operator in particular for molecules with a permanent dipole moment \cite{Resonances_Review_Domcke,Local_vs_Nonlocal_1981,DEA_HNCO_Dipole_Cizek,Star_Formation_H2_Cizek}. We expect that similar findings will be present for the light-induced resonances, but in contrast to the situation found for the natural resonances, the findings for the latter are controllable.   

Last but not least, we mention the arising overwhelming possibilities due to several atoms or molecules in the cavity.

\section*{Acknowledgments}
\noindent The authors thank A. I. Kuleff and H.-D. Meyer for valuable contributions. Financial support by the European Research Council (ERC) (Advanced Investigator Grant No. 692657) is gratefully acknowledged.

\bibliographystyle{apsrev}

%

\end{document}


\author{Lorenz S. Cederbaum}
\email{Lorenz.Cederbaum@pci.uni-heidelberg.de}
\affiliation{Theoretische Chemie, Physikalisch-Chemisches Institut, Universit\"at Heidelberg, Im Neuenheimer Feld 229, Heidelberg D-69120, Germany}
\author{Jacqueline Fedyk}
\affiliation{Theoretische Chemie, Physikalisch-Chemisches Institut, Universit\"at Heidelberg, Im Neuenheimer Feld 229, Heidelberg D-69120, Germany}

\title{Activating Cavity by Electrons \\
	 Supplementary Materials}

\date{\today}

\renewcommand{\thefigure}{S\arabic{figure}}

\maketitle
\section*{Contents}
 {\bf I. The resonant T-matrix and the total scattering state
 
 II. The computational model
 
 III. Resonance energies 
 
 IV. Results for resonances close to threshold}

\newpage

\section{The resonant T-matrix and the total scattering state} \label{The resonant T-matrix and the total scattering state}
The T-matrix describes a scattering process and can be used to calculate the respective cross section as well as the total scattering state \cite{Davydov_Book,Gottfried_Book,Lippmann_Schwinger_Weinberg_Book}. For Hamiltonians which can be written as $H=H_0+H_1$, where $H_0$ describes the incoming and outgoing scattering states and $H_1$ the interaction which gives rise to the scattering process, the T-matrix satisfies the following equation 
 
\begin{align}\label{T-matrix}
T = H_1 + H_1(E_I - H_0 +i0^+)^{-1}T,	
\end{align}

where $E_I$ is the initial energy of the scattering process and $0^+$ is a positive infinitesimal. The T-matrix can be expressed via the expansion

\begin{align}\label{T-matrix-expansion}
	T = H_1 + H_1(E_I - H_0 +i0^+)^{-1}H_1 + ... ,	
\end{align}

which can be utilized to compute its elements. 

In our case of an atom or molecule exposed to quantum light, the Hamiltonian reads \cite{Cohen_Tannoudji_Book,Feist_PhysRevX,Oriol_Cavity_CP} 

\begin{align}\label{Matter-Cavity-Hamiltonian}
	H = H_e + \hbar\omega_c\hat{a}^\dagger \hat{a} + g_0 \vec{e}_c\cdot \vec{d}(\hat{a}^\dagger + \hat{a}),
\end{align}

where the various terms are defined in the main text. Identifying the unperturbed Hamiltonian as the electronic part and the cavity photons of the set up with $H_0=H_e + \hbar\omega_c\hat{a}^\dagger \hat{a}$, and the interaction between them with $H_1=H_{mc}=g_0 \vec{e}_c\cdot \vec{d}(\hat{a}^\dagger + \hat{a})$, leads to

\begin{align}\label{T_Matrix}
	T = H_{mc} + H_{mc}(E_I - H_e - \hbar\omega_c\hat{a}^\dagger \hat{a} +i0^+)^{-1}T.
\end{align}

As discussed in the main text, initially, the target atom or molecule is in its ground electronic state $\Phi^{N}_{0}$ of energy $E^{N}_0$, the incoming electron of momentum $p$ has the kinetic energy $\epsilon_p=p^2/2m$, and the cavity is at rest and has zero photons. This initial state is denoted as $|I,p,0_c\rangle = |\Phi^{N+1}(p),0_c\rangle$ and its energy is $E_I = E^{N}_0 + \epsilon_p$. Other scattering states, where the electron has the energy $\epsilon_k$ are analogously denoted by $|\Phi^{N+1}(k),0_c\rangle$. The matter-cavity interaction $H_{mc}$ couples states of the same total number of electrons, but with a different number of cavity photons. As a result, a scattering state $|\Phi^{N+1}(k),0_c\rangle$ can couple to a completely different type of state where the $N+1$ electrons form a bound state $\Phi^{N+1}_{0}$ of energy $E^{N+1}_0$ and the cavity has one photon. This state is denoted by $|\Phi^{N+1}_{0},1_c\rangle$ and its energy is $E^{N+1}_0 + \hbar\omega_c$. The quantity $E^{N}_0-E^{N+1}_0$ is nothing but the electron affinity of the target. As discussed in the main text, the coupling between the two kinds of states, the scattering states $|\Phi^{N+1}(k),0_c\rangle$ and the bound state $|\Phi^{N+1}_{0},1_c\rangle$, can turn the latter state into a resonance which has a finite lifetime and decays into the electronic continuum. This resonance is a hybrid matter-light resonance and can be born only due to the presence of the quantum light. In the main text it has been called quantum light-induced resonance or briefly light-induced resonance and as nowadays quantum light is realized in a cavity, we may also call it cavity-induced resonance. 

In this work we are interested in the impact of this new kind of resonances on scattering and electron capture process and investigate the T-matrix for resonant scattering. We assume that the target atom or molecule itself does not possess intrinsic resonances (called natural resonances) in the relevant energy range in order to demonstrate this impact clearly. If the target does have natural resonances in the energy range investigated, the situation is extremely striking as there is a controlled interplay or competition between the induced and natural resonance. Controlled, because in contrast to the natural resonance whose properties are given by nature, we can change the properties of the induced resonance by varying the quantum light frequency $\omega_c$ and coupling strength $g_0$. The interplay between the induced and natural resonance is addressed in the main text but goes beyond the present context. To calculate the resonant scattering process, we utilize the expansion (\ref{T-matrix-expansion}) in the space of $|\Phi^{N+1}(k),0_c\rangle$ and $|\Phi^{N+1}_{0},1_c\rangle$ which are eigenstates of the unperturbed Hamiltonian $H_0$. 

We introduce the interaction matrix elements 
                          
\begin{align}\label{interaction-matrix-elements}
\gamma_{k0}
&=\langle\Phi^{N+1}(k),0_c|H_{mc}|\Phi^{N+1}_{0},1_c\rangle, \nonumber \\
&= g_0 d_{k0}, 	
\end{align}

and mention that they are linear functions of the coupling strength and the dipole transition moment $d_{k0}$ between the anionic state $\Phi^{N+1}_{0}$ and a continuum state $\Phi^{N+1}(k)$ with $k$ being parallel to the polarization direction $\vec{e}_c$ of the cavity mode. Applying the T-matrix to the initial state $|I,p,0_c\rangle = |\Phi^{N+1}(p),0_c\rangle$ also provides the total scattering state $|\Psi^+_p\rangle$ which is given by
 
\begin{align}\label{total-scattering-state}
	|\Psi^+_p\rangle
	&=|I,p,0_c\rangle + (E_I - H_e - \hbar\omega_c\hat{a}^\dagger \hat{a} +i0^+)^{-1}H_{mc}|\Psi^+_p\rangle, \nonumber \\
	&= |I,p,0_c\rangle + (E_I - H_e - \hbar\omega_c\hat{a}^\dagger \hat{a} +i0^+)^{-1}T|I,p,0_c\rangle, 	
\end{align}

and note that the expansion (\ref{T-matrix-expansion}) actually also provides the total scattering state. 

Employing the expansion one finds

\begin{align}\label{expansion-terms}
	T|I,p,0_c\rangle
	= & |\Phi^{N+1}_{0},1_c\rangle\gamma_{0p} + \sum_{k}|\Phi^{N+1}(k),0_c\rangle\gamma_{k0}(\epsilon_p - \epsilon_c +i0^+)^{-1}\gamma_{0p} + ..., \nonumber \\
	|\Psi^+_p\rangle 
	=& |I,p,0_c\rangle + |\Phi^{N+1}_{0},1_c\rangle(\epsilon_p - \epsilon_c +i0^+)^{-1}\gamma_{0p} \hspace{0.2cm} +    \nonumber  \\ 
	+ &  \sum_{k}|\Phi^{N+1}(k),0_c\rangle(\epsilon_p - \epsilon_k +i0^+)^{-1}\gamma_{k0}(\epsilon_p - \epsilon_c +i0^+)^{-1}\gamma_{0p} + ...,  
\end{align}

where $\epsilon_c=E^{N+1}_0-E^{N}_0+\hbar\omega_c$ and $E^{N}_0-E^{N+1}_0$ is the electron affinity of the target.

As shown in Refs.(\cite{Resonance_Vibrational_Structure_1977,Resonances_Review_Domcke}), these series can be summed up exactly. The result reads

\begin{align}\label{expansion-summed-up}
	T|I,p,0_c\rangle
	= & \sum_{k}|\Phi^{N+1}(k),0_c\rangle\gamma_{k0}(\epsilon_p - \epsilon_c - F(\epsilon_p))^{-1}\gamma_{0p}  +           \nonumber \\
	+ & |\Phi^{N+1}_{0},1_c\rangle F(\epsilon_p)(\epsilon_p - \epsilon_c - F(\epsilon_p))^{-1}\gamma_{0p} \nonumber \\
	|\Psi^+_p\rangle - |I,p,0_c\rangle
	= &   |\Phi^{N+1}_{0},1_c\rangle(\epsilon_p - \epsilon_c - F(\epsilon_p))^{-1}\gamma_{0p} \hspace{0.2cm} +    \nonumber  \\ 
	+ &  \sum_{k}|\Phi^{N+1}(k),0_c\rangle(\epsilon_p - \epsilon_k +i0^+)^{-1}\gamma_{k0}(\epsilon_p - \epsilon_c - F(\epsilon_p))^{-1}\gamma_{0p}, 	
\end{align}

where the complex shift function reads 

\begin{align}\label{Complex_Level_Shift}
F(\epsilon_p) = \Delta(\epsilon_p) - \frac{i}{2}\Gamma(\epsilon_p) = \sum_{k} \frac{|\gamma_{k0}|^2}{\epsilon_p - \epsilon_k +i0^+},
\end{align}

and will be evaluated numerically for an example in the next section where also the general relationship between the real and imaginary parts of the complex shift function is given.

\section{The computational model} \label{The computational model}
In the illustrative example chosen in the man text, the Hydrogen atom is the target for the impinging electron. We adopt the three dimensional potential (in atomic units) employed to compute the photodetachment of H$^-$ by strong high-frequency linearly polarized laser pulses \cite{Model_Potential_H_anion}

\begin{align}\label{Model_Potential}
V(r) = -V_0 e^{-r^2/r_0^2},
\end{align}  

where $V_0= 0.3831087$ and $r_0=2.5026$. The anion possesses a single bound state. The radial Schr\"odinger equation is solved numerically for both the angular momenta $l=0$ and $l=1$ in a radial box of length $L$ for a number of box radii by diagonalizing the Hamiltonian of the radial Schrödinger equation numerically to get the
corresponding eigenvalues and eigenvectors. The Hamiltonian matrix is written in a discrete variable
representation (DVR), which is a pseudo-spectral method \cite{DVR_1}. As basis functions we choose
the eigenfunctions of the quantum mechanical 1D “Particle in a box” problem, which are sine
functions. Their boundary conditions define the behavior of our computed eigenvectors, which are
consequently zero at the boundaries. In the sine-DVR, the grid points, the weights, the matrix
of the unitary transformation between the sine function basis and the grid basis, as well as the grid
basis can be calculated analytically. This allows one to directly built the potential energy matrix. The
evaluation of the effective potential at the calculated grid points, provides the elements of the diagonal
matrix. Getting the kinetic energy matrix is more tricky. After computing the matrix in the basis of
the sine functions, the before analytically calculated unitary transformation matrix is applied to
finally get the kinetic energy matrix in the grid basis.  

\subsection{Complex level shift function}
To determine the complex level shift function $F(\epsilon_p)$, we first calculate the transition dipoles between the ground state $\psi_0(l=0)=\sqrt{\dfrac{1}{4 \pi}} R(r)$ and all the computed states $\psi_k(l=1)= \sqrt{\dfrac{3}{4 \pi}} \cos(\vartheta) R_k(r)$ in the radial box, which we enumerate by $k$, along the polarization direction which is chosen to be the $z$ axis

\begin{align}\label{Transition_Dipoles}
	 \braket{\psi_0(l=0)|z|\psi_k(l=1)} = d_{0k},
\end{align}

and 

\begin{align}
	d_{k0}^* = d_{0k}.
\end{align}

We remind that the coupling elements are provided by $\gamma_{0k} = d_{0k}g_0$, where $g_0$ is the coupling strength of the cavity. 

To evaluate the complex level shift function in Eq.(\ref{Complex_Level_Shift}), we first make use of the relation \cite{Math_Phys_Book}
 
\begin{align}
\frac{1}{\epsilon_p - \epsilon_k +i0^+} = P\frac{1}{\epsilon_p - \epsilon_k} -i\pi\delta(\epsilon_p - \epsilon_k),
\end{align}

where P stands for the Cauchy principal value, and transform the sum into an integral \cite{Correlation_Effecs_Atoms_Kelly}

\begin{align}\label{sum_integral}
\sum_k \rightarrow \dfrac{L}{\pi} \int_0^{\infty} dk
\end{align}

for the radial momenta. Due to the appearance of the $\delta$-function, the integration to obtain the imaginary level shift can be made explicitly, giving

\begin{align}\label{Gamma}
\Gamma(\epsilon_p) = \pi \, \dfrac{8}{\sqrt{2}} \, L \, \dfrac{|\gamma_{p0}|^2}{\sqrt{\epsilon_p}} \,  .
\end{align}

The real level shift function can be obtained from the imaginary one, by employing the relationship \cite{Resonances_Review_Domcke}

\begin{align}\label{Delta}
\Delta(\epsilon_p) = \dfrac{1}{2\pi}\, P \, \int_0^{\infty} d\epsilon_k \, \dfrac{\Gamma(\epsilon_k)}{\epsilon_p - \epsilon_k},
\end{align}

and has to be evaluated numerically.

\begin{figure}[H]
	\begin{center}
		\includegraphics[width=10cm]{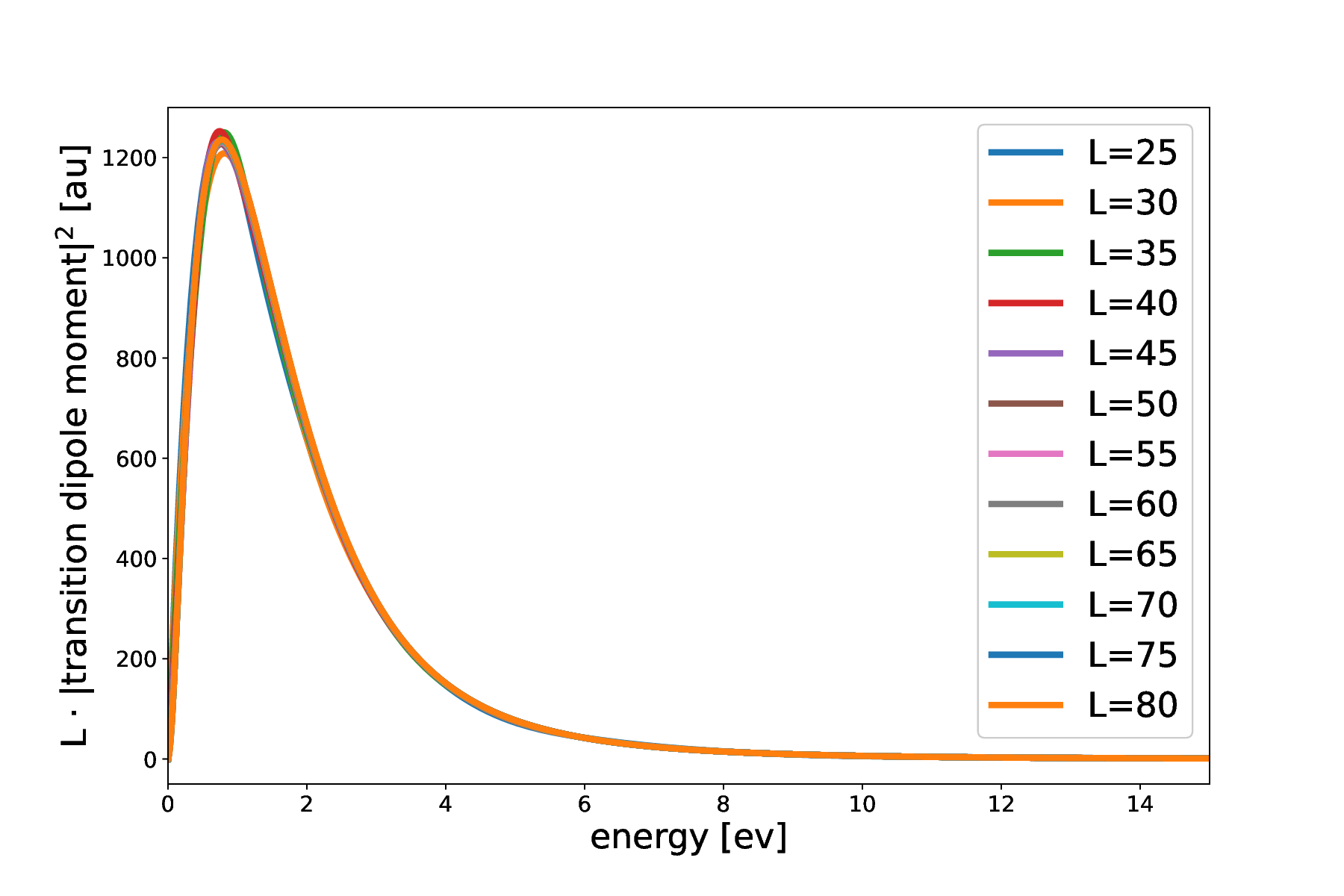}
		\caption{Demonstration of convergence achieved with respect to the box size $L$ used. The complex level shift function explicitly depends on the distribution of the absolute squares of the dipole transition moments multiplied by $L$, see Eqs. (\ref{Gamma},\ref{Delta},\ref{interaction-matrix-elements}). Shown are these distributions computed for several values of $L$.} 
		\label{fig:Transition_Dipoles_Convergence}
	\end{center}
\end{figure}

To demonstrate the convergence achieved with respect to the box size $L$ used, we show in Fig. (\ref{fig:Transition_Dipoles_Convergence}) the distribution of the absolute squares of the dipole transition moments computed for several box sizes. When multiplied by $L$, these distributions converge with growing box size to within the line width used to draw the curves. In all the results shown in the main text and below, the largest box size $L=80\,au$ is used.    

\subsection{Electron scattering cross section and electron capture probability}
Once the complex level shift function is calculated, the total resonant electron scattering cross section $\sigma (\epsilon_p)$ can be computed using \cite{Resonance_SCattering_N2_1}

\begin{align}\label{cross_section}
	\sigma (\epsilon_p) = \dfrac{2 \pi}{\epsilon_p} \, \dfrac{\Gamma^2(\epsilon_p)}{(\epsilon_p - \epsilon_c - \Delta(\epsilon_p))^2 + (\Gamma(\epsilon_p)/2)^2},
\end{align}

where the energy of the discrete state interacting with the continuum to become a resonance is, for electron-H scattering:

\begin{align}
	\epsilon_c = E_0^{N+1} - E_0^N + \hbar\omega_c = -0.754\,eV + \hbar\omega_c .
\end{align}

We now turn to the probability to capture an electron due to the presence of the cavity. For that purpose we inspect the weight of the hybrid captured electron state $|\Phi^{N+1}_{0},1_c\rangle$ in the total scattering state $|\Psi^+_p\rangle$. Using the exact expression for the latter in Eq.(\ref{expansion-summed-up}), one immediately finds

\begin{align}
|\langle \Phi^{N+1}_{0},1_c|\Psi^+_p\rangle|^2 =	\dfrac{|\gamma_{p0}|^2}{(\epsilon_p - \epsilon_c - \Delta(\epsilon_p))^2 + (\Gamma(\epsilon_p))^2/4}.
\end{align}

There is no meaning to probability at a point, and we have to sum over the states in a small energy interval $\delta_{\epsilon_p}$ around $\epsilon_p$. Within this interval the value of the complex level shift function does not change significantly, and we are left with a sum over $|\gamma_{p0}|^2$ only. Using the mapping (\ref{sum_integral}) and the relation (\ref{Gamma}) between the coupling elements and the width function $\Gamma(\epsilon_p)$, we find  

\begin{align}
\sum_p |\gamma_{p0}|^2 \rightarrow  \dfrac{1}{2\pi}\Gamma(\epsilon_p) \delta_{\epsilon_p},
\end{align}

where the sum is over states in the small energy interval $\delta_{\epsilon_p}$.

The capture probability is finally given by

\begin{align}\label{capture_probability}
	P_{capture} = \dfrac{1}{2 \pi} \dfrac{\Gamma(\epsilon_p) \delta_{\epsilon_p}}{(\epsilon_p - \epsilon_c - \Delta(\epsilon_p))^2 + (\Gamma(\epsilon_p))^2/4}.
\end{align}

It accounts for the probability to capture an electron impinging within a small energy interval $\delta_{\epsilon_p}$ around $\epsilon_p$. 

As discussed in the main text, cavities are subject to losses which lead to a finite lifetime of the cavity photon, see, e.g., Ref.~\cite{Cavity_Dynamics_Polaritons_Sfeir}. This decay of the cavity photon can be taken into account \cite{Cavity_Losses_Feist} by augmenting the cavity Hamiltonian $\hbar\omega_c\hat{a}^\dagger   \hat{a} \rightarrow \hbar\omega_c\hat{a}^\dagger \hat{a} -i\Gamma_{phot}/2$. Inspection of the T-matrix and total scattering state (\ref{expansion-summed-up}), shows how the cavity photon decay rate $\Gamma_{phot}$ enters the denominator in the cross section and in the capture probability, giving rise to the overall width $(\Gamma(\epsilon_p) + \Gamma_{phot})$ of the resonance.




\section{Resonance energies } \label{Resonance_energies}   
The structure of the T-Matrix, Eq. (\ref{expansion-summed-up}) is that of a discrete electronic state of energy $\epsilon_c$ embedded in the continuum and interacting with it \cite{Resonances_Review_Domcke}. In a cavity, the discrete state's energy is explicitly $\epsilon_c=E^{N+1}_0-E^{N}_0+\hbar\omega_c$ and describes the energy of the hybrid anion-light state $|\Phi^{N+1}_{0},1_c\rangle$ relative to the energy of the target. The energy of the discrete state is shifted by $\Delta(\epsilon_p)$ and the discrete state acquires a finite lifetime by $\Gamma(\epsilon_p)$ via its interaction with the continuum. It is common to define the resonance energy as the solution of \cite{Resonances_Review_Domcke}

\begin{align}\label{Resonance-energy}
E_{res} = \epsilon_c + \Delta(E_{res}).
\end{align}

I.e., where for $\epsilon_p=E_{res}$ the real part of the denominator of the T-matrix vanishes. Assuming that the imaginary part $\Gamma(\epsilon_p)/2$ varies smoothly with energy, the scattering cross section peaks at the resonance energy $\epsilon_p=E_{res}$. 

\begin{figure}[H]
	\begin{center}
		\includegraphics[width=10cm]{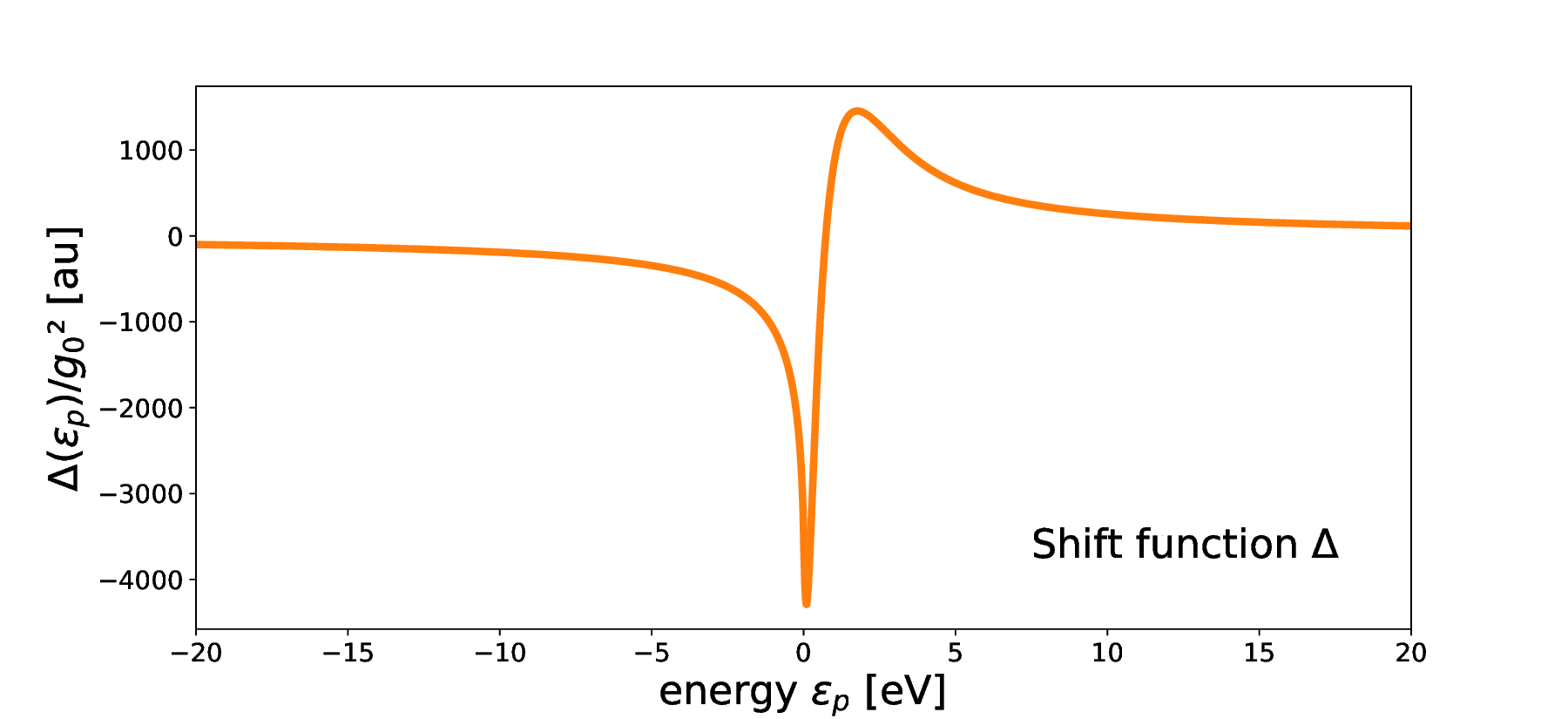}
		\caption{The entire level shift function $\Delta$. Note that while the width function $\Gamma$ vanishes for negative argument, this is not the case for $\Delta$. As a consequence, solutions of equation (\ref{Resonance-energy}) at negative energies are possible and give rise to bound states below the threshold of the continuum.}
		\label{fig:Delta_all_range}
	\end{center}
\end{figure}

While the width function $\Gamma(\epsilon_p)$ vanishes for negative argument, the shift function $\Delta(\epsilon_p)$ does not, see Eq. (\ref{Delta}). To have a complete picture of the situation of the resonance energy resulting from Eq. (\ref{Resonance-energy}), we depict the shift function in Fig. (\ref{fig:Delta_all_range}) also for negative energy values. We stress that solutions of Eq. (\ref{Resonance-energy}) at negative energies are meaningful and correspond to bound states. The coupling with the continuum of a discrete state lying in the continuum can give rise to a bound state below the threshold of the continuum if this coupling is sufficiently strong. Since the shift function $\Delta(\epsilon_p)$ depends quadratically on the cavity coupling strength $g_0$, one can control the strength of the coupling and thus the appearance of the structure of the cross section and capture probability, see also Section (\ref{Capture_probability_for_resonances_close_to_threshold}).
 
\begin{figure}[H]
	\begin{center}
				\includegraphics[width=10cm]{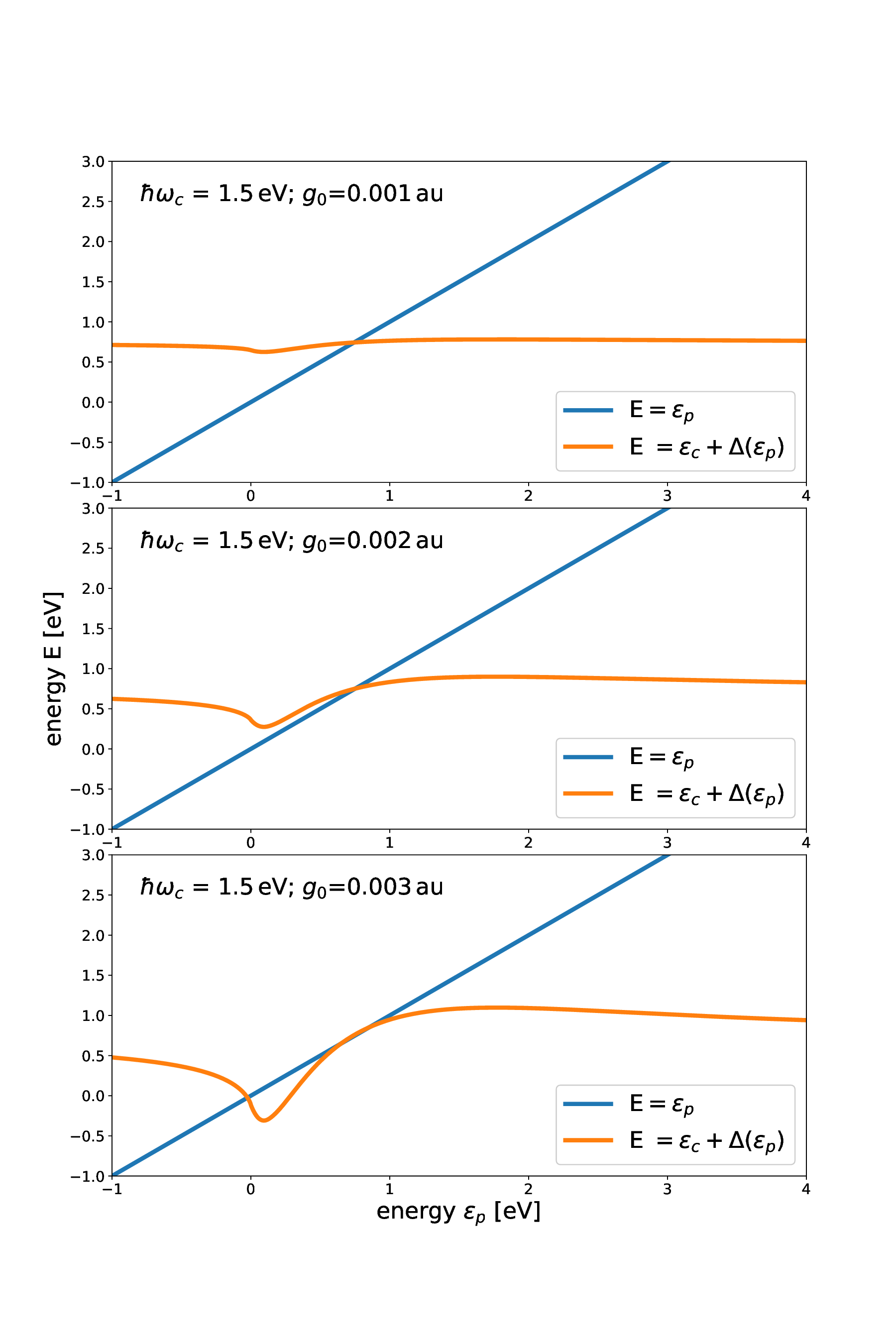}
		\caption{On the energy of the resonance for the H atom and $\hbar\omega_c=1.5$~eV. Shown are the two curves $E=\epsilon_p$ and $E=\epsilon_c + \Delta(\epsilon_p)$ as a function of energy $\epsilon_p$ for three values of the cavity coupling strength $g_0$. At their crossing point(s), the equation (\ref{Resonance-energy}) defining the resonance energy is fulfilled. See the text explaining the appearance of the respective scattering cross sections in the light of these curves.} 
		\label{fig:Resonance_energy_crossing_hw_c_1_5}
	\end{center}
\end{figure}

\begin{figure}[H]
	\begin{center}
		\includegraphics[width=10cm]{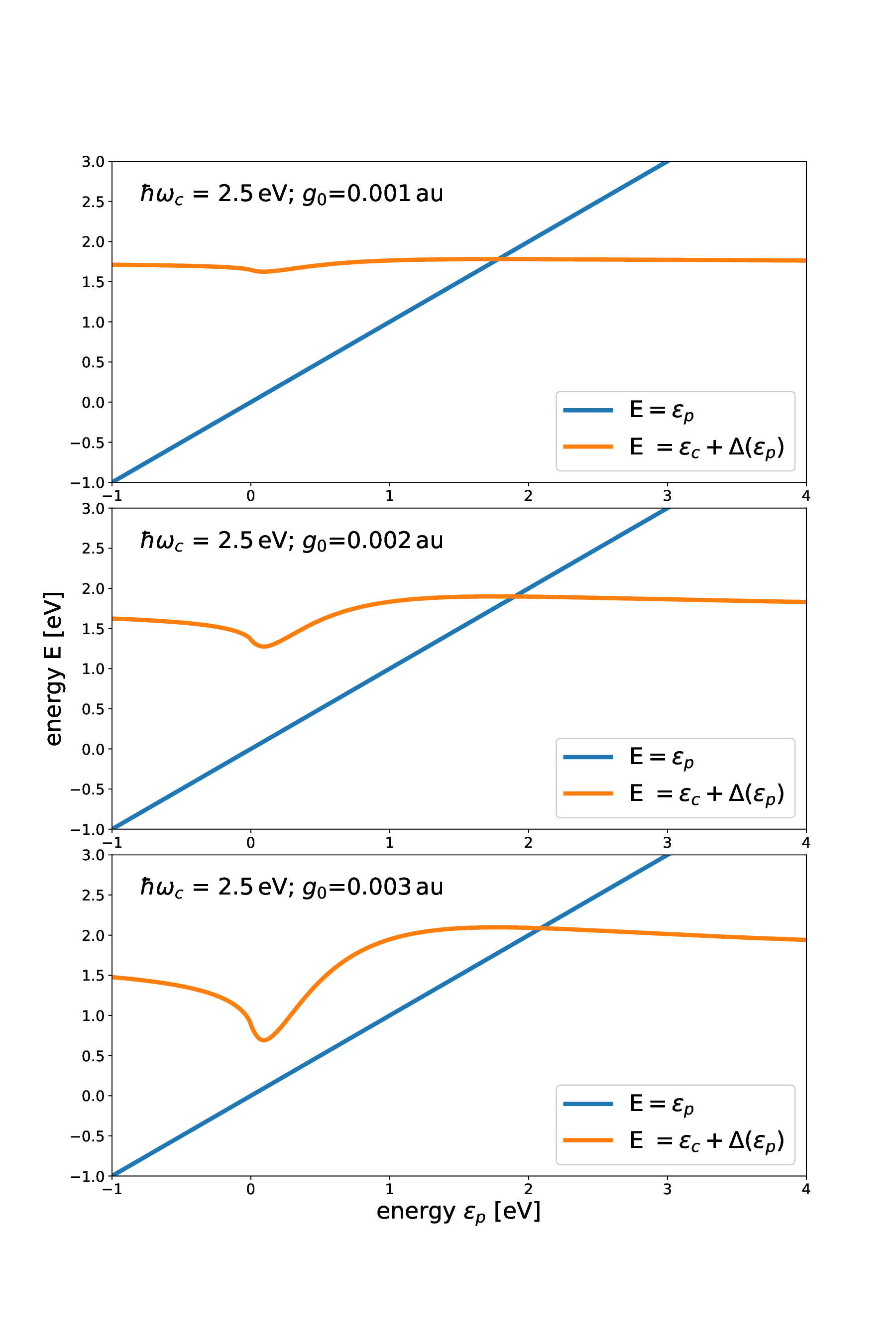}
		\caption{On the energy of the resonance for the H atom and $\hbar\omega_c=2.5$~eV. Shown are the two curves $E=\epsilon_p$ and $E=\epsilon_c + \Delta(\epsilon_p)$ as a function of energy $\epsilon_p$ for three values of the cavity coupling strength $g_0$. At their crossing point(s), the equation (\ref{Resonance-energy}) defining the resonance energy is fulfilled. See the text explaining the appearance of the respective scattering cross sections in the light of these curves.} 
		\label{fig:Resonance_energy_crossing_hw_c_2_5}
	\end{center}
\end{figure}

The solutions of Eq. (\ref{Resonance-energy}) for the scattering cross sections addressed in Fig. (3) of the main text are depicted in Figs. (\ref{fig:Resonance_energy_crossing_hw_c_1_5}) and (\ref{fig:Resonance_energy_crossing_hw_c_2_5}).
Each of these figures is for a given value of $\hbar\omega_c$ and shows three panels related to the different cavity coupling values of $g_0$ treated in the main text. Shown are the curves $E=\epsilon_p$ and $E=\epsilon_c + \Delta(\epsilon_p)$, the crossing points of which are solutions of Eq.~(\ref{Resonance-energy}). For the weak coupling case, the energy area around the crossing is smooth and one can anticipate a peaked structure of the cross sections at the energy of the resonance, symmetric for $\hbar\omega_c=2.5$~eV and less symmetric for $\hbar\omega_c=1.5$~eV due to the wiggle $\Delta(\epsilon_p)$ makes at low energy. For the other coupling strengths the situation is more interesting. For $\hbar\omega_c=2.5$~eV, the resonance energy is rather high and a peak can be seen in the cross section close to this energy, and due to the now more pronounced wiggle of $\Delta(\epsilon_p)$ and remembering that $\Gamma(\epsilon_p)$ is larger at lower energies, a second peak emerges in the cross sections at low energy (see Fig.~(3) of the main text). This lower lying peak is relatively more dominant for the strongest coupling strength, because the energy difference there between the two curves in the lower panel of Fig.~(\ref{fig:Resonance_energy_crossing_hw_c_2_5}) is clearly smaller than that in the middle panel. See Eq.~(\ref{cross_section}).    


A rather unusual situation occurs for the stronger couplings at $\hbar\omega_c=1.5$~eV. One can see in the middle panel of Fig. (\ref{fig:Resonance_energy_crossing_hw_c_1_5}) that the two curves $E=\epsilon_p$ and $E=\epsilon_c + \Delta(\epsilon_p)$ are very close to each other from about threshold to their crossing point. This explains the rather spread out structure of the cross section avoiding a two peak structure like that seen for $\hbar\omega_c=2.5$~eV in Fig.~(3) of the main text and discussed above. The situation is similar for the strongest studied coupling. Here, however, as seen in the lowest panel of Fig.~(\ref{fig:Resonance_energy_crossing_hw_c_1_5}), the two curves are not just close to each other over an extended energy range, but even exhibit more than a single crossing point. We may conclude that the cross sections, as well as the resonances can change substantially by varying the parameters of the quantum light used.

\newpage

\section{Results for resonances close to threshold} \label{Capture_probability_for_resonances_close_to_threshold}

In general, one can expect strong effects on the cross section if the resonance is close to threshold \cite{Resonances_Review_Domcke}. Whether a resonance of a free atom or molecule is close to threshold, is dictated by nature. In cavity, one can, in principle, tune the cavity such that the resonance gets close to threshold. Indeed, as the shift function varies substantially close to threshold, see Fig. (\ref{fig:Delta_all_range}), one may anticipate strong threshold effects as demonstrated in the main text.   

\begin{figure}[H]
	\begin{center}
		\includegraphics[width=10cm]{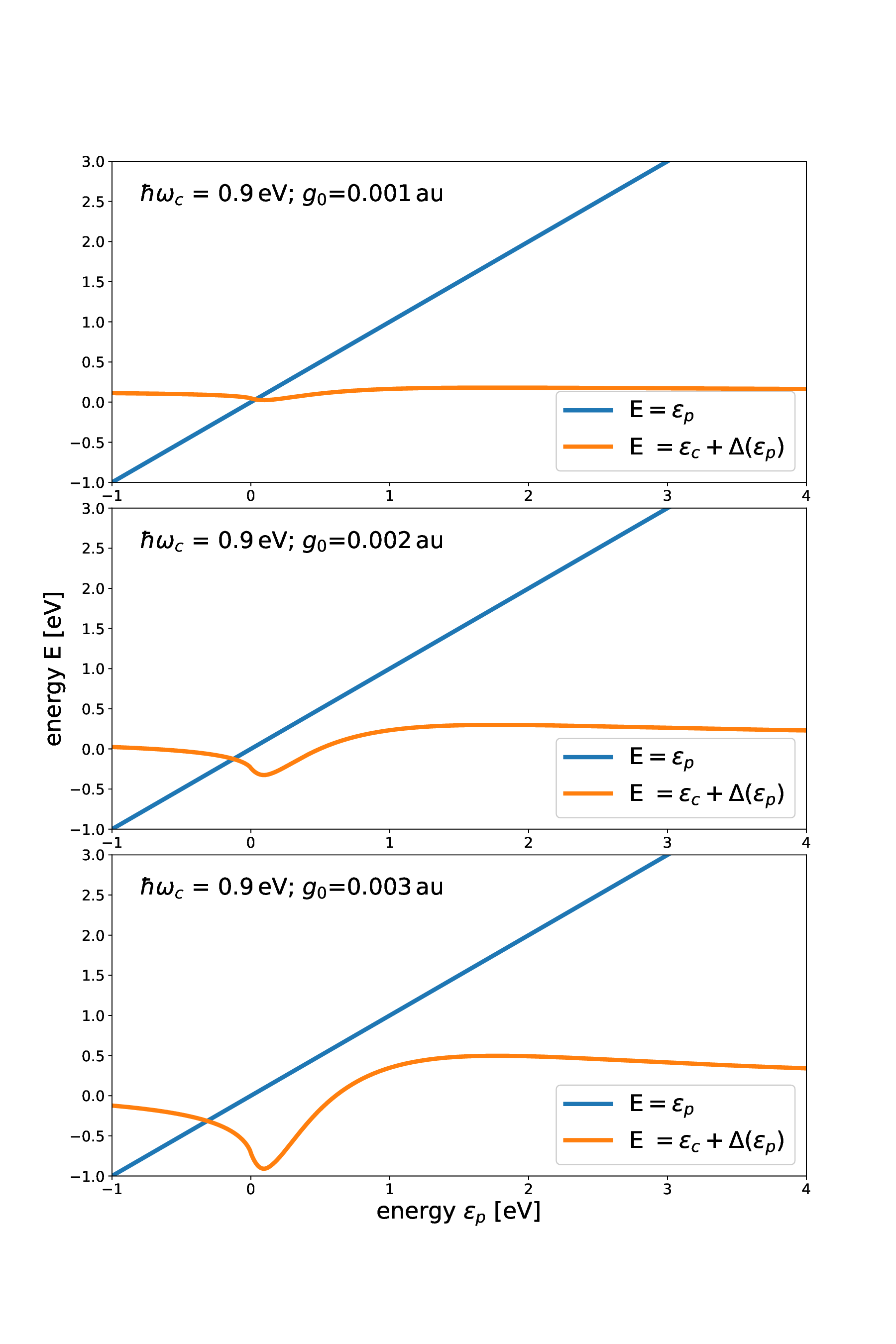}
		\caption{On the energy of the resonance for the H atom and $\hbar\omega_c=0.9$~eV. Shown are the two curves $E=\epsilon_p$ and $E=\epsilon_c + \Delta(\epsilon_p)$ as a function of energy $\epsilon_p$ for three values of the cavity coupling strength $g_0$. At their crossing point, the equation (\ref{Resonance-energy}) defining the resonance energy is fulfilled. For $\hbar\omega_c=0.9$~eV, the hybrid resonance for the smallest cavity coupling strength is close to threshold and for the stronger couplings it is pushed below threshold and becomes a hybrid bound state.} 
		\label{fig:Resonances_at_threshold}
	\end{center}
\end{figure}

Let us return to the example of the H atom and choose the cavity frequency such that for a given coupling strength the position of the emerging resonance is close to threshold. The calculated cross sections are collected in Fig. (5) of the main text for three values of the coupling strength. A resonance close to threshold has been shown there to lead to a high and peaked cross section close to threshold. In the upper panel, the photon energy $\hbar\omega_c=0.9$~eV creates a light-induced resonance close to threshold for the weakest coupling strength. As seen in the figure, the respective cross section strongly dominates those for the stronger coupling values. To better understand the finding that the weak coupling peak strongly dominates those for the stronger couplings, we compute the resonance positions for $\hbar\omega_c=0.9$~eV and all the three coupling strengths employed. Similarly as done above in Figs.~(\ref{fig:Resonance_energy_crossing_hw_c_1_5}) and (\ref{fig:Resonance_energy_crossing_hw_c_2_5}) for other values of $\hbar\omega_c$, we plot the two curves the two curves $E=\epsilon_p$ and $E=\epsilon_c + \Delta(\epsilon_p)$ and look for their crossing points which define the resonance energy position. The results for all three values of the cavity coupling strength are drawn in Fig.~(\ref{fig:Resonances_at_threshold}) for $\hbar\omega_c=0.9$~eV. It is seen in the upper panel that as expected, the resonance for the weakest coupling is close to threshold. Because the shift function $\Delta$ grows quadratically with the coupling strength, the respective solutions the equation $E_{res}=\epsilon_c + \Delta(E_{res})$ drop below the threshold and instead of a resonance, a bound hybrid state appears below threshold. The impact of these bound states on the respective cross sections is much weaker than that of a resonance close to threshold in spite of the larger coupling strengths. 

 The other panels of Fig.~(5) in the main text demonstrate that one can selectively enhance the cross section for a desired coupling strength. For any value of $g_0$, there is a value $\hbar\omega_c$ which creates a resonance close to threshold. At the same $\hbar\omega_c$, the resonance is pushed below the threshold and becomes a bound state for larger values of $g_0$, and pushed away from threshold and has thus typically less impact on the cross section for smaller values of $g_0$.

\begin{figure}[H]
	\begin{center}
		\includegraphics[width=8.5cm]{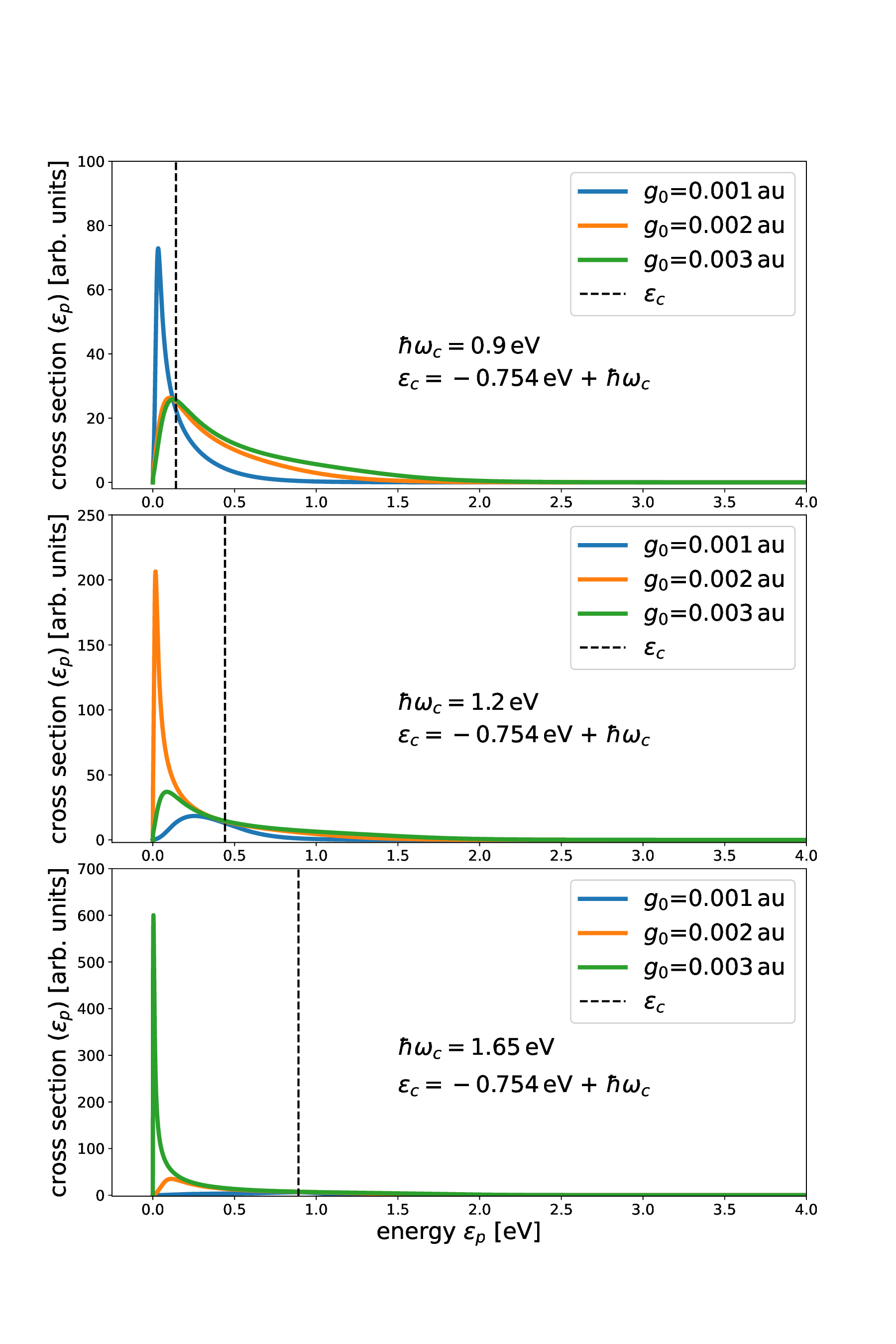}
		\caption{Illustrative examples of cross sections peaking close to threshold by controlling the cavity frequency for different coupling strengths. In contrast to the ideal cavity in the respective figure (5) of the main text, the present cavity is lossy. The lifetime of the cavity has been assumed to be $20$~fs, i.e., $\Gamma_{phot}=33$~meV. The energetic position of the discrete state embedded in and interacting with the continuum is drawn as a broken vertical line. Its energy depends linearly on the cavity frequency. This discrete state is shifted downwards in energy due to its interaction with the continuum and becomes a resonance due to its coupling to the continuum. As the shift depends quadratically on the coupling strength $g_0$, one can by choosing the photon energy $\hbar\omega_c$ control the location of the resonance. A resonance close to threshold leads to a high and peaked cross section. The photon energy $\hbar\omega_c=0.9$~eV brings the resonance for $g_0=0.001$~au close to threshold where it dominates the cross sections of the larger coupling strengths. In analogy, the other values of the photon energy shown are chosen to bring the resonance close to threshold for the other values of the coupling strength. Comparison with Fig.~(5) of the main text demonstrates that the measured cross sections reflect sensitively the cavity photon lifetime for resonances at threshold.} 
		\label{fig:cross_section_threshold_with_gamma_phot}
	\end{center}
\end{figure}

\newpage
Since nowadays cavities are lossy and the cavity photon has a finite lifetime, we have redone the calculations of the cross section with $\Gamma_{phot}=33$~meV, i.e., the lifetime of the cavity has been assumed to be $20$~fs. The results are collected in Fig.~(\ref{fig:cross_section_threshold_with_gamma_phot}). It is seen that as found for the situation without cavity losses, the cross section of the resonance near threshold strongly dominates the cross sections for the other coupling strengths at the same cavity frequency.  However, comparing with the results without losses shown in Fig.~(5) of the main text, one finds that there is a strong impact of the cavity lifetime on the cross sections of the resonances close to threshold! This is of particular interest. For many targets the electron affinity can be rather small compared to that of the H atom. Being close to threshold implies that one can use low frequency cavities for which there are cavities of longer lifetime \cite{Cavity_Review_Technology}. The measured cross sections reflect sensitively the cavity photon lifetime for resonances at threshold.

\begin{figure}[H]
	\begin{center}
		\includegraphics[width=10cm]{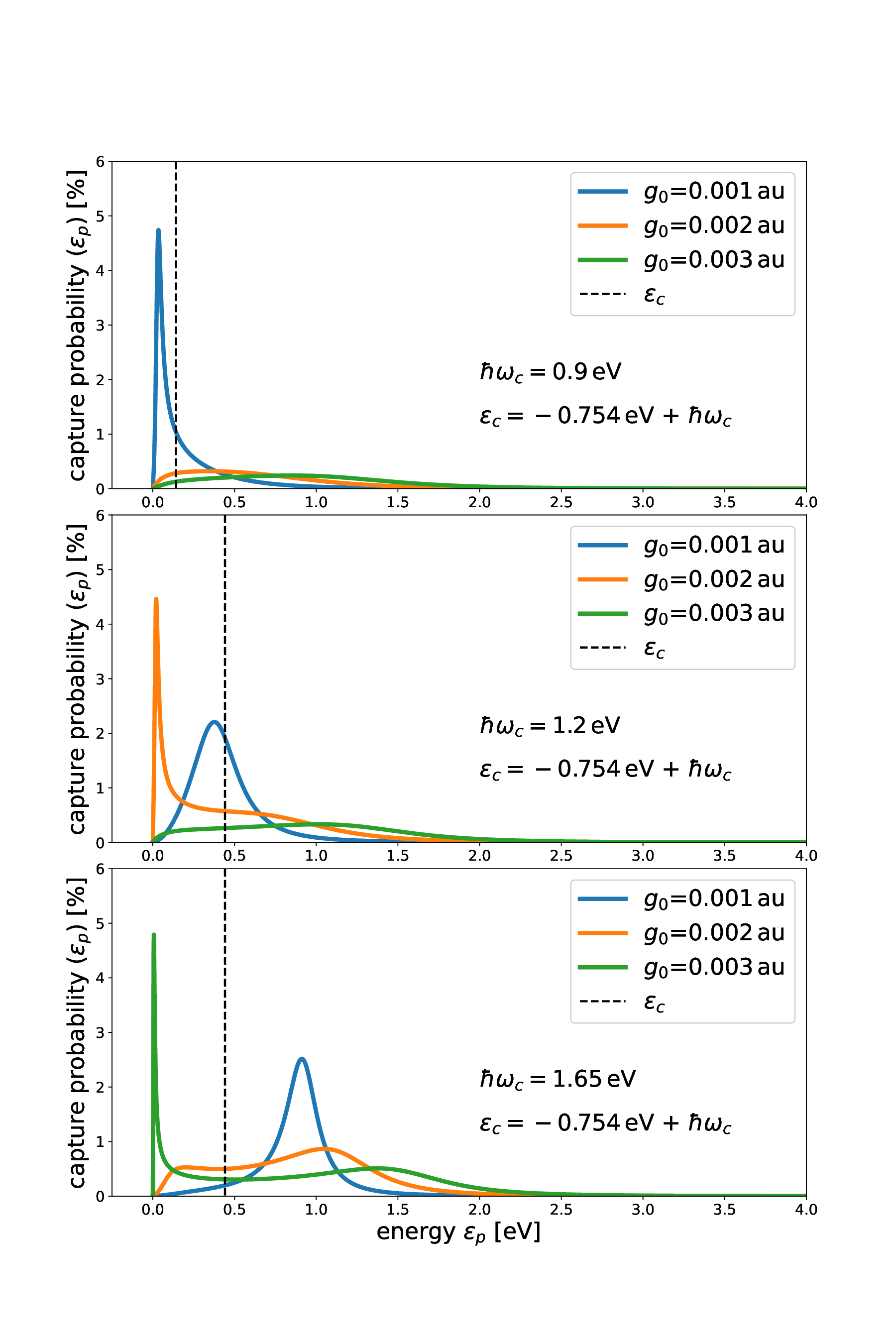}
		\caption{Capture probability for resonance close to threshold. Shown are the capture probabilities corresponding to the cross sections depicted in Fig. (\ref{fig:cross_section_threshold_with_gamma_phot}).} 
		\label{fig:Capture_probability_at_threshold}
	\end{center}
\end{figure}

To complement the discussion of the cross section of resonances close to threshold, we briefly discuss also the capture probability of the same resonances. The respective data is shown in Fig.~(\ref{fig:Capture_probability_at_threshold}) for the same conditions as for the cross sections shown in Fig.~(\ref{fig:cross_section_threshold_with_gamma_phot}). 

\begin{figure}[H]
	\begin{center}
		\includegraphics[width=10cm]{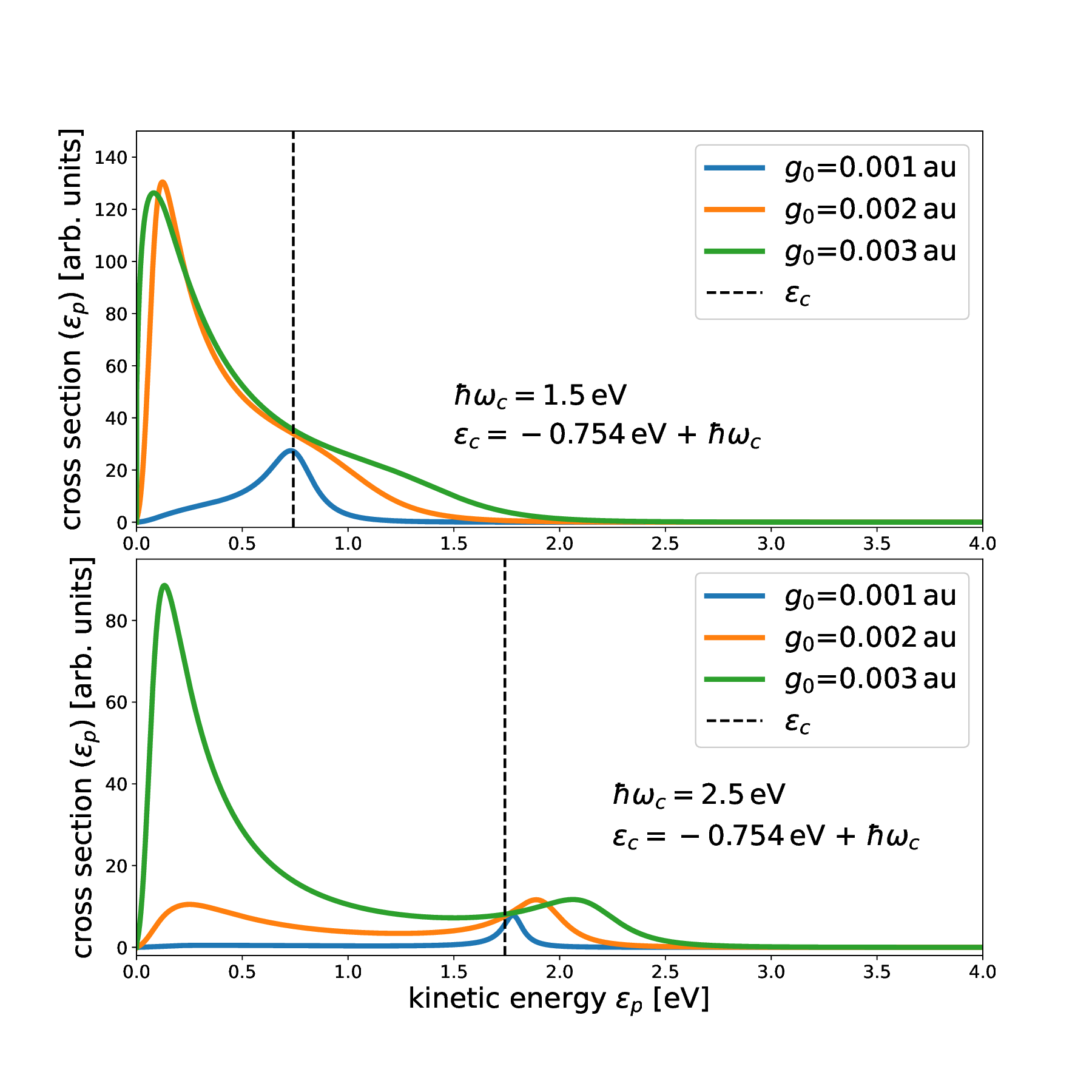}
		\caption{Illustrative examples of the electron scattering cross section for H atom in a cavity. Shown are cross sections for several values of the coupling strength $g_0$ and photon energy $\hbar\omega_c$ of the cavity. The energetic position of the discrete state embedded in and interacting with the continuum is drawn as a broken vertical line. Cavity losses (lifetime of the cavity of $20$~fs, i.e., $\Gamma_{photon}=33$~meV) have been taken into account. The results are to be compared with those shown in Fig.~(3) of the main text, where cavity losses have not been considered.} 
		\label{fig:cross_section_with_gamma_phot}
	\end{center}
\end{figure}

\begin{figure}[H]
	\begin{center}
		\includegraphics[width=10cm]{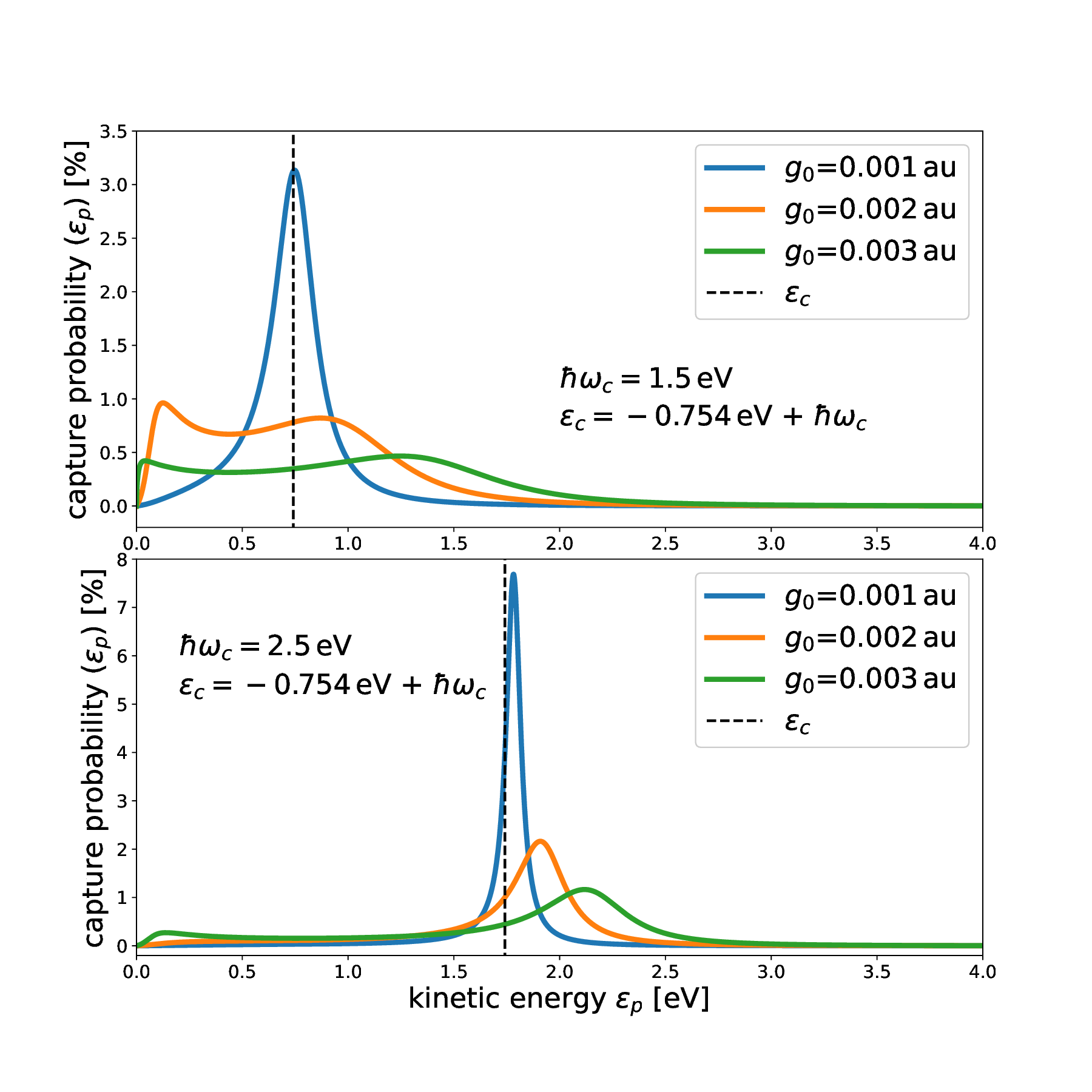}
		\caption{Illustrative examples of the electron capture probability by the H atom in a cavity. Shown are capture probabilities for several values of the coupling strength $g_0$ and photon energy $\hbar\omega_c$ of the cavity. No cavity losses are included. The results are to be compared with those shown in Fig. (4) of the main text, where cavity losses are taken into account leading to a lifetime of the cavity of $20$~fs, i.e., $\Gamma_{photon}=33$~meV. The probability relates to electrons in a kinetic energy interval of $\delta_{\epsilon_p}=0.01$~eV. The energetic position of the discrete state embedded in and interacting with the continuum is drawn as a broken vertical line.} 
		\label{fig:Capture_Probability_No_Photon}
	\end{center}
\end{figure}

For completeness we would like to show the impact of the finite lifetime on the cross sections for resonances far from threshold depicted in Fig.~(3) of the main text for the situation without losses, and, on the capture probability depicted in Fig.~(4) of the main text with losses. The cross sections computed including the losses can be seen in Fig.~(\ref{fig:cross_section_with_gamma_phot}). Comparing with Fig.~(3) of the main text shows some small impact of the losses, very negligible compared to the impact the losses have for resonances close threshold. The capture probabilities computed without taking losses into account are collected in Fig.~(\ref{fig:Capture_Probability_No_Photon}).   
 
 \newpage

\newpage


\bibliographystyle{apsrev}

%